\documentclass[a4paper,11pt]{article}
\pdfoutput=1 % if your are submitting a pdflatex (i.e. if you have
\usepackage{jheppub} % for details on the use of the package, please
\usepackage[T1]{fontenc} % if needed

\title{\boldmath 
Quantum entanglement and Bell nonlocality 
in top-quark pair production at a photon linear collider
}

\author[a,1]{Seong Youl Choi,\note{These authors contributed equally to this work
and share first authorship.}}
\author[a,1]{Dong Woo Kang,}
\author[b,1]{Jae Sik Lee}
\author[b,1]{and Chan Beom Park}

\affiliation[a]{Department of Physics, Jeonbuk National University, Jeonju 54896, Korea}
\affiliation[b]{Department of Physics and IUEP, Chonnam National University,
Gwangju 61186, Korea}

\emailAdd{sychoi@jbnu.ac.kr}
\emailAdd{dongwookang@jbnu.ac.kr}
\emailAdd{jslee@jnu.ac.kr}
\emailAdd{cbpark@jnu.ac.kr}

\abstract{
A photon linear collider, the two-photon collision mode of an $e^+e^-$ linear
collider,  uses high-energy laser photons backscattered off the incoming
electrons and positrons. The colliding-photon polarization is fully
controllable through the polarizations of the initial
electron and positron beams and laser photons.
We investigate the impact of colliding-photon polarization on the observability
of quantum entanglement in top-quark pair production
at a photon linear collider.
Constructing the spin density matrix of the $t\bar{t}$ two-qubit system from the
helicity amplitudes, we demonstrate that a photon linear collider is an ideal
machine to probe quantum entanglement
and Bell nonlocality across the broad phase space of the
process.
}

\newcommand{\imag}{\Im {\rm m}}
\newcommand{\real}{\Re {\rm e}}
\newcommand{\lsim}{\raisebox{-0.13cm}{~\shortstack{$<$ \\[-0.07cm] $\sim$}}~}
\newcommand{\gsim}{\raisebox{-0.13cm}{~\shortstack{$>$ \\[-0.07cm] $\sim$}}~}

\begin{document} 
{\small
\begin{flushright}
%  Last modified by CBP on \today \\
IUEP-HEP-26-01 \\[1mm]
%(Dated: {\color{red}\today})
(July 31, 2026)
\end{flushright} }
\vspace{-1.5cm}
\maketitle
\flushbottom

%\newpage
%------------------------------------------------------------------
\section{Introduction}
\label{sec:introduction}
At colliders, the {\it measurement} takes place when the polarized particles
decay and quantum entanglement is accessible through
correlations among particle spins~\cite{Barr:2024djo}.
The short lifetime of top quark, $\tau_t \sim 10^{-25}$~s, is much shorter than
the spin-correlation timescale of $\sim 10^{-21}$~s~\cite{Bigi:1986jk},
making it a good probe of quantum entanglement at colliders,
owing to the simplicity and analyticity of the two-qubit quantum state.
Quantum entanglement and Bell inequality violation in top-quark pair production
at the LHC were first proposed in Refs.~\cite{Afik:2020onf,Fabbrichesi:2021npl}.
Indeed, the ATLAS and CMS collaborations at the LHC have reported the
observation of quantum entanglement with a significance exceeding
$5\sigma$ near the $t\bar t$ production threshold~\cite{ATLAS:2023fsd,CMS:2024pts}
and at high $t\bar t$ invariant masses~\cite{CMS:2024zkc}.

Quantum entanglement gives rise to the stronger spin correlations,
than those expected in the classical theory,
among the states which once interacted but, being separated spatially,
cannot interact now.  At colliders, the spin correlations are accessible
through the distribution of the momenta of
the final state into which the original particle decays
\cite{Baumgart:2012ay}.
The possibilities to probe quantum entanglement at colliders were first suggested in
Refs.~\cite{Tornqvist:1980af,Tornqvist:1986pe,Privitera:1991nz,Abel:1992kz}.
More recently, it was shown that quantum entanglement is experimentally accessible
in the Higgs boson decays into charged gauge bosons~\cite{Barr:2021zcp},
using two-qubit systems such as top-quark, tau-lepton and photon pairs
\cite{Fabbrichesi:2022ovb,Dong:2023xiw},
in the Higgs boson decay into tau leptons~\cite{Altakach:2022ywa},
in weak gauge boson production~\cite{Fabbrichesi:2023cev},
and in $B$ meson decays~\cite{Fabbrichesi:2023idl}.
Since the ATLAS and CMS collaborations reported
observation of quantum entanglement at the LHC,
several proposals have appeared to measure
quantum entanglement and Bell nonlocality at the
LHC~\cite{Zhang:2025mmm,Fabbrichesi:2025psr,Goncalves:2025mvl,Ruzi:2025jql,
Goncalves:2025xer,DelGratta:2025xjp,Flacke:2025dwk,Antozzi:2026vdi},%
\footnote{
% It is interesting to note that
%Note that the formation of
The formation of toponium at the LHC
\cite{Fadin:1990wx,CMS:2025kzt,ATLAS:2025kvb,Fuks:2025wtq} 
could be involved in the
analysis of quantum entanglement near the $t\bar t$ production
threshold~\cite{Flacke:2025dwk,Antozzi:2026vdi}.}
an $e^+e^-$ collider
\cite{Maltoni:2024csn,Fabbrichesi:2024wcd,Cheng:2024btk,
Han:2025ewp,Altakach:2026fpl,Guo:2026yhz,
Gabrielli:2026tnl, Zhang:2026nwm, Yang:2026uwu},
and a muon collider~\cite{Ding:2025mzj}.
Other proposals target
the Higgs boson decays into four fermions via two vector
bosons~\cite{DelGratta:2025qyp} and
the decay of neutral kaons into two and three
pions~\cite{Fabbrichesi:2025zpw}.
For the current status of quantum entanglement and Bell nonlocality
at colliders, we refer to
Refs.~\cite{Barr:2024djo,Fabbrichesi:2025aqp,Afik:2025ejh}.

In this work, we investigate quantum entanglement and Bell nonlocality
in top-quark pair production at a photon linear collider (PLC),
$\gamma\gamma \to t\bar{t}$.
A PLC is the two-photon collision mode of an
$e^+e^-$ linear collider, using backscattered laser photons off
the incoming electrons and positrons~\cite{Ginzburg:1981vm, Ginzburg:1982yr,
Kuhn:1992fx,Bigi:1992pr,ECFADESYPhotonColliderWorkingGroup:2001ikq,
Asner:2001vh,Barklow:2023ess,
%Berger:2024ora,
LinearColliderVision:2025hlt,LinearCollider:2025lya}.
The capability of controlling the colliding-photon polarizations,
combined with the clean experimental environment,
makes a PLC an ideal machine to probe
quantum entanglement and Bell inequality violation.

This paper is organized as follows.
In Section~\ref{sec:QE}, we introduce the spin density matrix
of a two-qubit system composed of two spin-$1/2$ particles
and briefly review the four criteria for quantum entanglement and
Bell inequality violation considered in this work.
Section~\ref{sec:SPD} is devoted to developing the spin density matrix
for a two-qubit system constructed from the helicity amplitudes.
The formalism is process- and model-independent and
classifies the polarization vectors and spin correlations
according to their P, CP, and CP$\widetilde{\mathrm{T}}$ parities.
In Section~\ref{sec:LwSPD}, we apply the formalism to construct
the luminosity-weighted spin density matrix for polarized
colliding photons.
Through numerical analysis in Section~\ref{sec:Numerical}, we show that the
capability of controlling the colliding-photon polarizations at a PLC
significantly enhances the observability of quantum entanglement and Bell
inequality violation.
Section~\ref{sec:Conclusions} contains our conclusions and summary.
In Appendix~\ref{sec:16Cs}, we present the 16 polarization
coefficients introduced in Section~\ref{sec:SPD}.
%

%\newpage

%------------------------------------------------------------------
\section{Quantum entanglement of a two-qubit system}
\label{sec:QE}

The present work examines the bipartite system of a top and anti-top quark
pair explicitly. The framework described in this section, however, is
applicable to any bipartite
system $S_A \otimes S_B$ of two spin-$1/2$ particles, where $S_A$ and $S_B$ represent
the individual spin-$1/2$ systems. The quantum state of such a system can be represented
by the spin density matrix of a two-qubit
system:
\begin{eqnarray}
\label{eq:rho_0}
\rho=\frac{1}{4}\left[
{\bf 1}_2\otimes{\bf 1}_2 +
\sum_{i=1}^3 B_i^+ \left(\sigma_i\otimes{\bf 1}_2\right) +
\sum_{j=1}^3 B_j^- \left({\bf 1}_2\otimes\sigma_j\right) +
\sum_{i,j=1}^3 C_{ij} \left(\sigma_i\otimes\sigma_j\right) \right],
\end{eqnarray}
where ${\bf 1}_2$ is the $2 \times 2$ identity matrix and $\sigma_{1,2,3}$ are the three
Pauli matrices, defined as
\begin{eqnarray}
{\bf 1}_2=\left(\begin{array}{cc} 1 & 0 \\ 0 & 1 \end{array}\right)\,; \
\sigma_1=\left(\begin{array}{cc} 0 & 1 \\ 1 & 0 \end{array}\right)\,, \
\sigma_2=\left(\begin{array}{cc} 0 & -i \\ i & 0 \end{array}\right)\,, \
\sigma_3=\left(\begin{array}{cc} 1 & 0 \\ 0 & -1 \end{array}\right)\,.
\end{eqnarray}
The polarizations of the two particles are represented by the real coefficients obtained
by taking the traces
\begin{eqnarray}
B_i^+={\rm Tr}[\rho \left(\sigma_i\otimes{\bf 1}_2\right)]\,, \
B_j^-={\rm Tr}[\rho \left({\bf 1}_2\otimes\sigma_j\right)]\,,
\end{eqnarray}
and their spin correlations are given by
\footnote{The following identities hold for tensor products:
$(A \otimes B)(C \otimes D) = AC \otimes BD$ and
${\rm Tr}(A \otimes B) = {\rm Tr}(A) {\rm Tr}(B)$.}
\begin{eqnarray}
C_{ij}={\rm Tr}[\rho\left(\sigma_i\otimes\sigma_j\right)]\,,
\end{eqnarray}
where the indices $i$ and $j$, ranging from $1$ to $3$, refer to the top and anti-top
quarks, respectively, in the subsequent concrete analysis. Explicitly, the spin
density matrix $\rho$ in Eq.~\eqref{eq:rho_0} can be expressed in a $4\times 4$
matrix form as
%
%\begin{small}
%\begin{footnotesize}
%\begin{scriptsize}
%\begin{tiny}
\begin{equation}
\label{eq:rho_gen}
\hskip -0.3cm
\begin{scriptsize}
%\rho\!=\! 
\rho= 
\frac{1}{4}\!\left(\begin{array}{cc|cc}
 1+B^{+}_3+B^{-}_{3}+C_{33} &
 \! B^{-}_1+C_{31}-i(B^{-}_2+C_{32})\! &
 B^{+}_1+C_{13}-i(B^{+}_2+C_{23}) &
 \! C_{11}-C_{22}-i(C_{12}+C_{21}) \\
 B^{-}_1+C_{31}+i(B^{-}_2+C_{32}) &
 1+B^{+}_3-B^{-}_{3}-C_{33} &
 \! C_{11}+C_{22}+i(C_{12}-C_{21}) &
 B^{+}_1-C_{13}-i(B^{+}_2-C_{23}) \\ \hline
 B^{+}_1+C_{13}+i(B^{+}_{2}+C_{23}) &
 \! C_{11}+C_{22}+i(C_{21}-C_{12})\! &
 1-B^{+}_3+B^{-}_{3}-C_{33} &
 B^{-}_1-C_{31}-i(B^{-}_{2}-C_{32})\\
 C_{11}-C_{22}+i(C_{21}+C_{12}) &
 \! B^{+}_1-C_{13}+i(B^{+}_2-C_{23})\! &
 B^{-}_{1}-C_{31}+i(B^{-}_{2}-C_{32}) &
 1-B^{+}_3-B^{-}_{3}+C_{33}\\[2mm]
\end{array} \!\right).
\end{scriptsize}
\end{equation}
%\end{small}
%\end{footnotesize}
%\end{scriptsize}
%\end{tiny}
%
This explicit form is particularly useful for analyzing
the quantum entanglement of the bipartite system.

For a bipartite system of two spin-$1/2$ particles with spin density matrix $\rho$,
we adopt the following four established and complementary criteria for quantum entanglement and Bell inequality violation:
\begin{itemize}
\item[-]\underline{Peres-Horodecki criterion}~\cite{Peres:1996dw}:
\begin{equation}
\label{eq:PH}
\rho^{T_B}=
(\mathbf{1}_A \otimes T_B)[\rho]<\,0
~ \Longleftrightarrow ~ \ {\rm entangled} \,,
\end{equation}
where $T_B$ denotes partial transposition on the subsystem $S_B$ and the
inequality $\rho^{T_B}<\,0$ means that at least one of the four eigenvalues of
the partially transposed matrix $\rho^{T_B}$ is negative.
The Peres-Horodecki criterion is comprehensive and thorough. It is applicable
not only to two qubits ($d = d_A = d_B = 2$) but also to a qubit-qutrit system
($d_A = 2$, $d_B = 3$).
To quantify the Peres-Horodecki criterion, we use the negativity, ${\cal N}[\rho]$, as an entanglement quantifier, defined as
\begin{equation}
\label{eq:Negativity}
{\cal N}[\rho]=\sum_k\frac{|\lambda_k|-\lambda_k}{2}=
\sum_{\lambda_k<0} |\lambda_k|\,,
\end{equation}
where $\lambda_k$ are the eigenvalues of the partially transposed density matrix $\rho^{T_B}$.
\item[-]\underline{Concurrence}~\cite{Wootters:1997id}:
\begin{equation}
\label{eq:Concurrence}
{\cal C}[\rho]={\rm max}\left(0,r_1-r_2-r_3-r_4\right)  > 0
~ \Longleftrightarrow ~ \ {\rm entangled}\,,
\end{equation}
where $r_{1,2,3,4}$ with $r_1\geq r_{2,3,4}$
are the {\it square roots} of nonnegative eigenvalues of the $4\times 4$
auxiliary matrix $R$ constructed using
a two-qubit $4\times 4$ density matrix $\rho$:
% \footnote{Note that $R$ is not Hermitian but its eigenvalues are non-negative.}
%
\begin{equation}
R = \rho\,\left(\sigma_2\otimes\sigma_2\right)\,
\rho^*\,\left(\sigma_2\otimes\sigma_2\right)\,,
\end{equation}
which is non-Hermitian but its eigenvalues are always nonnegative.
\item[-]\underline{Clauser-Horne-Shimony-Holt (CHSH) inequality
augmented by the Horodecki condition}~\cite{Clauser:1969ny,Horodecki:1995nsk}:
\begin{equation}
\label{eq:CHSH}
m_{12}\equiv m_1 + m_2 >1
~ \Longleftrightarrow ~ \ {\rm violation~of~Bell~inequality}\,,
\end{equation}
where $m_1\geq m_2\geq m_3$ are the three eigenvalues of the following symmetric
matrix
\begin{equation}
M \ = \ CC^T\,,
\end{equation}
with $C$ being the $3\times 3 $ spin correlation matrix denoted by
$C_{ij}$ in Eq.~\eqref{eq:rho_0}.
Note that
the violation of the Bell inequality is a sufficient condition
for quantum entanglement.
\item[-]\underline{Entanglement marker $D$}:
\begin{equation}
\label{eq:D}
D \ \equiv \ \frac{1}{3}\,\Delta_E < -\frac{1}{3}
~ \Longrightarrow ~ \ {\rm entangled}\,,
\end{equation}
where~\cite{Afik:2020onf}
\begin{eqnarray}
\label{eq:DE}
\Delta_E= C_{11}- | C_{22}+C_{33} |  \,.
\end{eqnarray}
%
% While this criterion is only a sufficient condition for quantum entanglement derived from the Peres-Horodecki
% criterion, it provides a highly reliable means of detecting quantum entanglement.
While this criterion is only a sufficient condition for quantum entanglement
derived from the Peres-Horodecki criterion, it provides a practical means of
detecting quantum entanglement.
When the diagonal components of the $3\times 3$ correlation matrix $C$ are all
negative, $D = {\rm Tr}[C]/3$, which is the observable adopted by the ATLAS and
CMS collaborations to probe entanglement near the $t\bar{t}$ production
threshold~\cite{ATLAS:2023fsd,CMS:2024pts}.
\end{itemize}
%

%------------------------------------------------------------------
\section{Spin density matrix of a two-qubit system}
\label{sec:SPD}

\begin{figure}[t!]
%\begin{figure}[h!]
%\begin{figure}[b!]
\vspace{-0.5cm}
\begin{center}
\includegraphics[width=14.0cm]{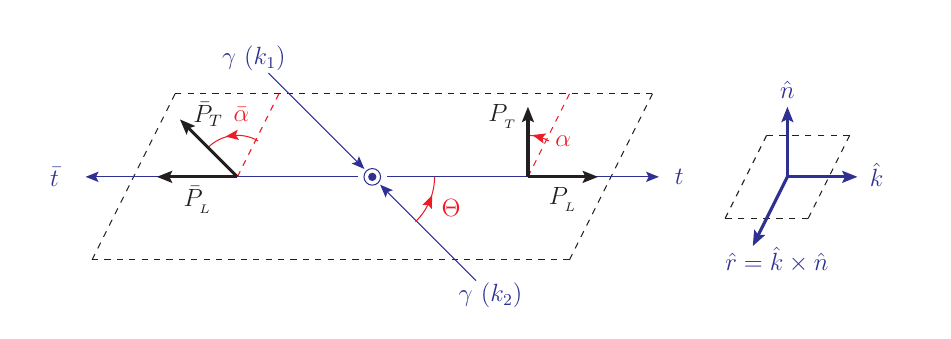}
\end{center}

\caption{\it
The $t\bar t$ production plane with the definition
of the scattering angle $\Theta$ for the process
$\gamma(k_1, \lambda_1)\, +\, \gamma(k_2,\lambda_2)
\to t(p,\sigma)\, + \, \bar{t}(\bar{p}, \bar{\sigma})$.
The four-momentum $k_1$ ($k_2$) is assigned to the photon
backscattered off the incoming $e^-$ ($e^+$) at the PLC.
The transverse polarization vectors $P_T$ and $\bar{P}_T$
have azimuthal angles $\alpha$ and $\bar{\alpha}$
appearing in Eq.~\eqref{eq:PDM_ttbar},
respectively, measured  from the production plane.
The $\{\hat n\,,\hat r\,,\hat k\}$ basis is defined by
$\hat k = {\vec p}/|{\vec p}\,|$,
$\hat n = {\vec k}_1\times{\vec p}/|{\vec k}_1\times{\vec p}\,|$, and
$\hat r = \hat k\times\hat n$.
Note that, in the $\{\hat n\,,\hat r\,,\hat k\}$ basis,
$P_L = P_k$,
$P_T c_\alpha = -P_r$, and
$P_T s_\alpha = P_n$ for the top quark, whereas for the anti-top quark,
$\bar P_L = - \bar P_k$,
$\bar P_T c_{\bar\alpha} = -\bar P_r$, and
$\bar P_T s_{\bar\alpha} = \bar P_n$.
}
\label{fig:helicity_basis}
\end{figure}

The two-photon process $\gamma\gamma \to t\bar{t}$ with on-shell top and
anti-top quarks receives two tree-level contributions.  For the process
\begin{eqnarray}
\label{eq:Process}
\gamma(k_1, \lambda_1)\, +\, \gamma(k_2,\lambda_2)
\ \ \to \ \ t(p,\sigma)\, + \, \bar{t}(\bar{p}, \bar{\sigma})\,,
\end{eqnarray}
where $\lambda_1$ and $\lambda_2$ denote the helicities of the colliding photons
and $\sigma$ and $\bar{\sigma}$ denote the helicities of the top and anti-top quarks,
the helicity amplitude in the $\gamma\gamma$ center-of-mass frame is given
by~\cite{Asakawa:2000jy}
\begin{eqnarray}
\label{eq:HelAmp}
{\cal M}_{\lambda_1\lambda_2;\sigma\bar\sigma}\equiv {\cal A}_C\,
\langle \sigma\bar\sigma;\lambda_1\lambda_2\rangle =
\frac{8\pi\alpha Q_t^2}{1-\beta^2 \cos^2\Theta} \,
\langle \sigma\bar\sigma;\lambda_1\lambda_2\rangle
\ \ \mbox{with} \ \
{\cal A}_C = \frac{8\pi\alpha Q_t^2}{1-\beta^2 \cos^2\Theta}\,,
\end{eqnarray}
where $\lambda_1\,,\lambda_2\,, \sigma\,, \bar{\sigma}=\pm$ are used for
notational convenience.
Here, $Q_t = 2/3$, $\beta^2 = 1 - 4M_t^2/\hat{s}$, and $\Theta$ is the angle between
$\vec{k}_1$ and $\vec{p}$ in the production plane, with $\hat{s} = (k_1 + k_2)^2 = (p
+ \bar{p})^2$. The photon backscattered off the incoming $e^-$ ($e^+$) carries
four-momentum $k_1$ ($k_2$), as illustrated in Fig.~\ref{fig:helicity_basis}.
The reduced amplitudes $\langle \sigma\bar\sigma;\lambda_1\lambda_2\rangle$ are
given in Table~\ref{tab:aatt_SM}. 
Notably, the amplitude vanishes when
$\lambda_1 = \lambda_2$ and $\sigma = -\bar{\sigma}$ owing to exact
cancellations between the $t$- and $u$-channel diagrams.

\begin{table}[!hbt]
%\begin{table}[!t]
%\caption{\label{tab:aatt_SM} The reduced amplitudes
%$\langle \sigma\bar\sigma;\lambda_1\lambda_2\rangle$ in Eq.~\eqref{eq:HelAmp}.
%Here, $\beta = \sqrt{1 - 4M_t^2/\hat{s}}$, where $M_t$ denotes the pole mass of
%the top quark, and $\Theta$ is the angle between $\vec{k}_1$ and $\vec{p}$
%in the production plane (see Fig.~\ref{fig:helicity_basis}). }
\setlength{\tabcolsep}{0.5ex}
\renewcommand{\arraystretch}{1.4}
\begin{center}
\begin{tabular}{|c|cc|cc|}
\hline
$(\lambda_1\lambda_2)\downarrow ~ (\sigma\bar\sigma)\rightarrow$
& $\langle ++\rangle$ &  $\langle --\rangle$ &  $\langle -+\rangle$ &  $\langle +-\rangle$ \\
\hline
$\langle ++\rangle$ & $\frac{2M_t}{\sqrt{\hat s}}\,\big[1+\beta\big]$ &
$\frac{2M_t}{\sqrt{\hat s}}\,\big[1-\beta\big]$
&  0 &  0 \\
$\langle --\rangle$ & $\frac{2M_t}{\sqrt{\hat s}}\,\big[-1+\beta\big]$ &
$\frac{2M_t}{\sqrt{\hat s}}\,\big[-1-\beta\big]$
&  0 &  0 \\
\hline
$\langle -+\rangle$ & $-\frac{2M_t}{\sqrt{\hat s}}\beta\, \sin^2\Theta$
&  $\frac{2M_t}{\sqrt{\hat s}}\beta\, \sin^2\Theta$
&  $-\beta \sin\Theta\,(\cos\Theta+1)$ &  $-\beta \sin\Theta\,(\cos\Theta-1)$ \\
$\langle +-\rangle$ & $-\frac{2M_t}{\sqrt{\hat s}}\beta\,\sin^2\Theta$
&  $\frac{2M_t}{\sqrt{\hat s}}\beta \,\sin^2\Theta$
&  $-\beta \sin\Theta\,(\cos\Theta-1)$ &  $-\beta \sin\Theta\,(\cos\Theta+1)$ \\
\hline
\end{tabular}
\end{center}
\caption{\label{tab:aatt_SM} The reduced amplitudes
$\langle \sigma\bar\sigma;\lambda_1\lambda_2\rangle$ in Eq.~\eqref{eq:HelAmp}.
Here, $\beta = \sqrt{1 - 4M_t^2/\hat{s}}$, where $M_t$ denotes the pole mass of
the top quark, and $\Theta$ is the angle between $\vec{k}_1$ and $\vec{p}$
in the production plane (see Fig.~\ref{fig:helicity_basis}). }
\end{table}

Note that the reduced amplitudes transform under
P, CP, and CP$\widetilde{\rm T}$  as follows:
\begin{eqnarray}
\label{eq:CPTtransformations}
\langle \sigma\bar{\sigma};\lambda_1,\lambda_2\rangle
  &\stackrel{\rm P}{\longleftrightarrow} &
  - (-1)^{(\sigma-\bar{\sigma})/2}
  \,\langle -\sigma,-\bar\sigma;-\lambda_1,-\lambda_2\rangle \,;
\nonumber \\[2mm]
\langle \sigma\bar{\sigma};\lambda_1,\lambda_2\rangle
  &\stackrel{\rm CP}{\longleftrightarrow} &
  - (-1)^{(\sigma-\bar{\sigma})/2}
  \,\langle -\bar{\sigma},-\sigma;-\lambda_2,-\lambda_1\rangle \,;
\nonumber \\[2mm]
\langle \sigma\bar{\sigma};\lambda_1,\lambda_2\rangle
  &\stackrel{\rm CP\widetilde{\rm T}}{\longleftrightarrow} &
  - (-1)^{(\sigma-\bar{\sigma})/2}
  \,\langle -\bar{\sigma},-\sigma;-\lambda_2,-\lambda_1\rangle^* \,,
\end{eqnarray}
or equivalently,
\begin{eqnarray}
\langle \pm\pm;\lambda_1,\lambda_2\rangle
  &\stackrel{\rm P}{\longleftrightarrow} &
-\langle \mp\mp;-\lambda_1,-\lambda_2\rangle\,,\ \ \ \
\langle \pm\mp;\lambda_1,\lambda_2\rangle
  \stackrel{\rm P}{\longleftrightarrow}
\langle \mp\pm;-\lambda_1,-\lambda_2\rangle\,; \nonumber \\[2mm]
\langle \pm\pm;\lambda_1,\lambda_2\rangle
  &\stackrel{\rm CP}{\longleftrightarrow} &
-\langle \mp\mp;-\lambda_2,-\lambda_1\rangle\,,\ \ \ \
\langle \pm\mp;\lambda_1,\lambda_2\rangle
  \stackrel{\rm CP}{\longleftrightarrow}
\langle \pm\mp;-\lambda_2,-\lambda_1\rangle\,; \nonumber \\[2mm]
\langle \pm\pm;\lambda_1,\lambda_2\rangle
  &\stackrel{\rm CP\widetilde{\rm T}}{\longleftrightarrow} &
-\langle \mp\mp;-\lambda_2,-\lambda_1\rangle^*\,,\ \ \ \
\langle \pm\mp;\lambda_1,\lambda_2\rangle
  \stackrel{\rm CP\widetilde{\rm T}}{\longleftrightarrow}
\langle \pm\mp;-\lambda_2,-\lambda_1\rangle^*\,.
\end{eqnarray}
As a result of these transformation properties of the helicity amplitudes,
each polarization coefficient possesses definite P, CP, and
CP$\widetilde{\rm T}$ parities, as will be shown shortly.

To streamline the analysis of the $t\bar t$ spin correlations, we arrange the
helicity amplitudes in Eq.~\eqref{eq:HelAmp} into a $2\times 2$ matrix:
\begin{eqnarray}
{\cal M} = {\cal A}_C\left(\begin{array}{cc}
\langle ++;\lambda_1\lambda_2\rangle & \langle +-;\lambda_1\lambda_2\rangle \\
\langle -+;\lambda_1\lambda_2\rangle & \langle --;\lambda_1\lambda_2\rangle
\end{array}\right) \ \equiv \ {\cal A}_C\,{\bf M}\,.
\end{eqnarray}
Contracting the matrix ${\cal M}$ and its Hermitian conjugate ${\cal M}^\dagger$
with the $2\times 2$ spin density matrices for the top and anti-top quarks,
defined as
\begin{eqnarray}
\label{eq:PDM_ttbar}
\rho_t & = & \frac{1}{2}\left(\begin{array}{cc}
     1+P_L                  & P_T\,{\rm e}^{-i\alpha}  \\[1mm]
    P_T\,{\rm e}^{i\alpha} & 1-P_L
                   \end{array}\right)\,, \ \ \
\bar{\rho}_t = \frac{1}{2}\left(\begin{array}{cc}
     1+\bar{P}_L       & -\bar{P}_T\,{\rm e}^{i\bar{\alpha}}  \\[1mm]
    -\bar{P}_T\,{\rm e}^{-i\bar{\alpha}} & 1-\bar{P}_L
                   \end{array}\right)\,,
\end{eqnarray}
we obtain the polarization-weighted amplitude squared given
by~\cite{Hagiwara:1985yu}
\begin{equation}
\overline{\left|{\cal M}\right|^2}
\ =\ \frac{1}{4}\, \sum_{\lambda_1,\lambda_2=\pm}
{\rm Tr}\left[{\cal M}\bar{\rho}_t^T {\cal M}^\dagger\rho_t\right]
\ =\ \frac{1}{4}\,|{\cal A}_C|^2 \sum_{\lambda_1,\lambda_2=\pm}
{\rm Tr}\left[{\bf M}\bar{\rho}_t^T {\bf M}^\dagger\rho_t\right]
\ \equiv\ \frac{1}{4}\,|{\cal A}_C|^2\,\widehat\Sigma\,.
\end{equation}
By identifying
$P_L = P_k$, $P_T c_\alpha = -P_r$, and
$P_T s_\alpha = P_n$ for the top quark and
$\bar P_L = - \bar P_k$,
$\bar P_T c_{\bar\alpha} = -\bar P_r$ and
$\bar P_T s_{\bar\alpha} = \bar P_n$ for the anti-top quark
in the $\{\hat n\,,\hat r\,,\hat k\}$ basis,
see Fig.~\ref{fig:helicity_basis}, and evaluating
${\rm Tr}\left[{\bf M}\bar{\rho}_t^T {\bf M}^\dagger\rho_t\right]$
explicitly, we can obtain $\widehat\Sigma$ in terms of
the 16 polarization coefficients from $\widehat C_1$ to
$\widehat C_{16}$ as follows:
\begin{footnotesize}
\begin{eqnarray}
\label{eq:polarization_weighted_square}
\widehat\Sigma
&=& \big(\widehat C_1[+++]+\widehat C_3[+++]\big)
\nonumber \\[1mm]
&+&
P_n \big(\widehat C_6[++-]+\widehat C_8[+-+]\big)
+ P_r \big(-\widehat C_5[-++]-\widehat C_7[---]\big)
+ P_k \big(\widehat C_2[---]+\widehat C_4[-++]\big)
\nonumber \\[1mm]
&+&
 \bar P_n \big(\widehat C_6[++-]-\widehat C_8[+-+]\big)
+\bar P_r \big(-\widehat C_5[-++]+\widehat C_7[---]\big)
+\bar P_k \big(-\widehat C_2[---]+\widehat C_4[-++]\big)
\nonumber \\[1mm]
&+&
P_n\bar P_n \big(\widehat C_{13}[+++]-\widehat C_{15}[+++]\big)
+P_n \bar P_r \big(-\widehat C_{14}[--+]-\widehat C_{16}[-+-]\big)
+P_n\bar P_k \big(-\widehat C_{10}[--+]-\widehat C_{12}[-+-]\big)
\nonumber \\[1mm]
&+&
 P_r \bar P_n \big(\widehat C_{14}[--+]-\widehat C_{16}[-+-]\big)
+P_r\bar P_r \big(\widehat C_{13}[+++]+\widehat C_{15}[+++]\big)
+ P_r\bar P_k \big(\widehat C_9[+--]+\widehat C_{11}[+++]\big)
\nonumber \\[1mm]
&+&
 P_k \bar P_n \big(\widehat C_{10}[--+]-\widehat C_{12}[-+-]\big)
+ P_k \bar P_r \big(-\widehat C_9[+--]+\widehat C_{11}[+++]\big)
+P_k\bar P_k \big(-\widehat C_1[+++]+\widehat C_3[+++]\big)\,,
\nonumber \\
\end{eqnarray}
\end{footnotesize}
where the P, CP, and CP$\widetilde{\rm T}$ parities of each polarization
coefficient, derived from the transformation rules of the reduced amplitudes
given in Eq.~\eqref{eq:CPTtransformations}, are denoted in square brackets.
The explicit forms of all 16 polarization coefficients in terms
of the reduced amplitudes are given in Appendix~\ref{sec:16Cs}.
Our classification and parity assignments are in excellent agreement
with those reported in Ref.~\cite{Bernreuther:2015yna}.
%
%
%\vspace{-0.5cm}
\begin{table}[!hbt]
%\begin{table}[!t]
%\begin{table}[!b]
%
%\begin{footnotesize}
\begin{scriptsize}
%
%\caption{\label{tab:Parities_16Cs}
%The P, CP, and CP$\widetilde{\rm T}$ parities of the 16 polarization coefficients
%and their related physical quantities.}
\vspace{-0.5cm}
\setlength{\tabcolsep}{0.3ex}
\renewcommand{\arraystretch}{1.6}
\begin{center}
\begin{tabular}{|c|cccc|cccc|}
\hline
Polarization coefficient &
$\widehat C_1$ & $\widehat C_2$ & $\widehat C_3$ & $\widehat C_4$ &
$\widehat C_5$ & $\widehat C_6$ & $\widehat C_7$ & $\widehat C_8$ \\
\hline
[P, CP, CP$\widetilde{\rm T}$] Parities &
$[+++]$ & $[---]$ & $[+++]$ & $[-++]$ &
$[-++]$ & $[++-]$ & $[---]$ & $[+-+]$ \\[1mm]
Physical quantity &
$\dfrac{\widehat A-\widehat C_{kk}}{2}$ & $\dfrac{\widehat B_k^+ - \widehat B_k^-}{2}$ &
$\dfrac{\widehat A+\widehat C_{kk}}{2}$ & $\dfrac{\widehat B_k^+ + \widehat B_k^-}{2}$ &
$-\dfrac{\widehat B_r^+ + \widehat B_r^-}{2}$ & $\dfrac{\widehat B_n^+ + \widehat B_n^-}{2}$ &
$-\dfrac{\widehat B_r^+ - \widehat B_r^-}{2}$ & $\dfrac{\widehat B_n^+ - \widehat B_n^-}{2}$ \\[2mm]
\hline 
\hline
Polarization coefficient &
$\widehat C_9$ & $\widehat C_{10}$ & $\widehat C_{11}$ & $\widehat C_{12}$ &
$\widehat C_{13}$ & $\widehat C_{14}$ & $\widehat C_{15}$ & $\widehat C_{16}$  \\
\hline
[P, CP, CP$\widetilde{\rm T}$] Parities &
$[+--]$ & $[--+]$ & $[+++]$ & $[-+-]$ &
$[+++]$ & $[--+]$ & $[+++]$ & $[-+-]$ \\[1mm]
Physical quantity &
$\dfrac{\widehat C_{rk}-\widehat C_{kr}}{2}$ &
$\dfrac{\widehat C_{kn}-\widehat C_{nk}}{2}$ &
$\dfrac{\widehat C_{rk}+\widehat C_{kr}}{2}$ &
$-\dfrac{\widehat C_{kn}+\widehat C_{nk}}{2}$ &
$\dfrac{\widehat C_{nn}+\widehat C_{rr}}{2}$ &
$\dfrac{\widehat C_{rn}-\widehat C_{nr}}{2}$ &
$\dfrac{\widehat C_{rr}-\widehat C_{nn}}{2}$ &
$-\dfrac{\widehat C_{rn}+\widehat C_{nr}}{2}$ \\[2mm]
\hline
\end{tabular}
\end{center}
\caption{\label{tab:Parities_16Cs}
The P, CP, and CP$\widetilde{\rm T}$ parities of the 16 polarization coefficients
and their related physical quantities.}
%
%\end{footnotesize}
\end{scriptsize}
\end{table}

Table~\ref{tab:Parities_16Cs} lists the P, CP, and CP$\widetilde{\mathrm{T}}$
parities of the 16 polarization coefficients and the related physical
observables extractable from $t\bar{t}$ production and the angular
distributions of their decay products.
At leading order (LO) in QED, only the $[+++]$ parity coefficients $\widehat C_{1}$,
$\widehat C_{3}$, $\widehat C_{11}$, $\widehat C_{13}$, and $\widehat C_{15}$ are
nonzero. Consequently, all polarization vectors $\widehat B_i^+$ and
$\widehat B_j^-$ vanish.  The spin correlation matrix is then simplified, with
only the elements $\widehat C_{nn}$, $\widehat C_{rr}$, $\widehat C_{kk}$,
$\widehat C_{rk}$, and $\widehat C_{kr}$ remaining. Since
$\widehat C_{9}$ (with $[+--]$ parity) vanishes at LO,
we further have $\widehat C_{rk}=\widehat C_{kr}$.

While the preceding simplification holds at LO in QED,
it breaks down beyond LO, even within the Standard Model
(SM)~\cite{Bernreuther:2015yna}.  Electroweak and QCD quantum corrections,
incorporating P and CP violation in the electroweak sector, generate
absorptive parts and introduce P, CP, and CP$\widetilde{\rm T}$-violating
effects. This implies that all 16 polarization coefficients can potentially
contribute beyond LO and/or beyond the SM.  Importantly, our amplitude-based
formalism for constructing the spin density matrix of a two-qubit system
offers a crucial advantage. Its process- and model-independent structure,
combined with its ability to classify physical observables by their P, CP,
and CP$\widetilde{\rm T}$ parities, makes it a powerful probe for pinpointing
new-physics contributions beyond the SM, even in the presence of these complex
corrections.

The final form of $\widehat\Sigma$ in Eq.~\eqref{eq:polarization_weighted_square}
leads to the spin density matrix in the $\{\hat n\,,\hat r\,,\hat k\}$ basis,
\begin{small}
\begin{equation}
\label{eq:SDM_hat}
\rho=\frac{1}{4}\left[
{\bf 1}_2\otimes{\bf 1}_2 +
\frac{1}{\widehat A}\,
\sum_{i=n,r,k} \widehat B_i^+ \left(\sigma_i\otimes{\bf 1}_2\right) +
\frac{1}{\widehat A}\,
\sum_{j=n,r,k} \widehat B_j^- \left({\bf 1}_2\otimes\sigma_j\right) +
\frac{1}{\widehat A}\,
\sum_{i,j=n,r,k} \widehat C_{ij} \left(\sigma_i\otimes\sigma_j\right) \right]\,,
\end{equation}
\end{small}
where the
{\it unnormalized} polarization vectors
$\widehat B_i^+$ and $\widehat B_j^-$
for the top and anti-top quarks are given by
\begin{small}
\begin{eqnarray}
\label{eq:B_hat}
\widehat B_i^+&=&
\big(\widehat C_6[++-]+\widehat C_8[+-+],
-\widehat C_5[-++]-\widehat C_7[---],
\widehat C_2[---]+\widehat C_4[-++]\big)\,,\nonumber \\[2mm]
\widehat B_j^-&=&
\big(\widehat C_6[++-]-\widehat C_8[+-+],
-\widehat C_5[-++]+\widehat C_7[---],
-\widehat C_2[---]+\widehat C_4[-++]\big)\,,
\end{eqnarray}
\end{small}
respectively, and the {\it unnormalized} $3\times 3$ spin correlation matrix
$\widehat C_{ij}$ by
\begin{small}
\begin{eqnarray}
\label{eq:C_hat}
\widehat C_{ij}&=&
\left(\begin{array}{ccc}
\widehat C_{13}[+++]-\widehat C_{15}[+++] &
-\widehat C_{14}[--+]-\widehat C_{16}[-+-] &
-\widehat C_{10}[--+]-\widehat C_{12}[-+-] \\[1mm]
\widehat C_{14}[--+]-\widehat C_{16}[-+-] &
\widehat C_{13}[+++]+\widehat C_{15}[+++] &
\widehat C_9[+--]+\widehat C_{11}[+++] \\[1mm]
\widehat C_{10}[--+]-\widehat C_{12}[-+-] &
-\widehat C_9[+--]+\widehat C_{11}[+++] &
-\widehat C_1[+++]+\widehat C_3[+++]
\end{array}\right) \,. \nonumber \\
\end{eqnarray}
\end{small}
On the other hand, the normalization factor, defined by
\begin{eqnarray}
\label{eq:A_hat}
\widehat A = \widehat C_1[+++]+\widehat C_3[+++]\,,
\end{eqnarray}
is related to the unpolarized differential cross section.  
The explicit form of this relationship is
\begin{eqnarray}
\label{eq:cx_0}
{\rm d}\hat\sigma_0=
\frac{1}{2\hat s}\,N_c\,\overline{\sum|{\cal M}|^2} \,{\rm d}\Phi_2  
&=&
\frac{1}{2\hat s}\,N_c\,
\left[\frac{|{\cal A}_C|^2}{4}
\sum_{\lambda_1,\lambda_2,\sigma,\bar\sigma}
|\langle\sigma\bar\sigma;\lambda_1\lambda_2\rangle|^2\right] \,
\frac{\beta}{16\pi} \,{\rm d}\cos\Theta \nonumber \\
&=&\frac{\beta\,N_c}{32\pi\hat s}\,\left|{\cal A}_C\right|^2\,
\widehat{A}\,{\rm d}\cos\Theta\,,
\end{eqnarray}
with the color factor $N_c = 3$ in the SM.

In summary, we have developed an amplitude-level formalism for
constructing the spin density matrix of a two-qubit system.
While the present analysis focuses on the specific SM process
$\gamma\gamma \to t\bar t$ at a PLC, the formalism itself is not restricted to
this particular example. Its process- and model-independent structure provides
a systematic framework for classifying the polarization vectors and
spin correlations of the spin density matrix according to their fundamental
P, CP, and CP$\widetilde{\rm T}$ parities, making it applicable to a wide
range of processes and models beyond the SM.

%------------------------------------------------------------------
\section{Luminosity-weighted spin density matrix}
\label{sec:LwSPD}

A key advantage of PLCs is the efficient control of colliding photon helicities
through careful optimization of the initial electron and positron beam
polarizations and the laser beams used in the Compton backscattering
process~\cite{Ginzburg:1981vm, Ginzburg:1982yr,Kuhn:1992fx}. Beyond this efficient polarization
control, specific polarization configurations offer the remarkable possibility
of significantly enhancing the $\gamma\gamma$ luminosity within narrow regions
of $\sqrt{\hat s}$. These regions are of particular interest, as they may maximize
quantum entanglement and/or enable observation of Bell inequality violation.
%
% The polarizations of the initial electron and positron
% beams are denoted by $P_e$ and $\tilde P_e$ , respectively.
% For initial laser photons, $P_c$ and $\tilde P_c$ are degrees of
% circular polarization.
The electron and positron beam polarizations are denoted $P_e$ and
$\tilde{P}_e$, respectively; the degrees of circular polarization
of the initial laser photons are denoted $P_c$ and $\tilde{P}_c$.
Specifically,
the combinations $P_eP_c=\tilde P_e\tilde P_c=-1$, chosen for our numerical
analysis in the next section, exhibit such enhanced luminosity.
For this reason,
transverse polarizations of the laser beams are not considered
in the present study. In this section, we demonstrate how a PLC can be used
to achieve these enhanced capabilities.

From the Compton backscattering of laser photons, the luminosity of a PLC
is given by~\cite{Ginzburg:1982yr}
\begin{equation}
\frac{{\rm d}^2{\cal L}_{\gamma\gamma}^{\lambda_1\lambda_2}}
{{\rm d}y_1{\rm d}y_2} \ = \ 
\left(\frac{1}{\sigma_c^{\gamma_1}}\frac{{\rm d}\sigma_c^{\gamma_1}}{{\rm d}y_1}\right) \
\left(\frac{1}{\sigma_c^{\gamma_2}}\frac{{\rm d}\sigma_c^{\gamma_2}}{{\rm d}y_2}\right) \
{\cal L}_{ee}\,,
\end{equation}
where $y_i=E_{\gamma_i}/E_b$ denotes the ratio of the energy 
$E_{\gamma_i}$ of the colliding photon with the helicity $\lambda_{i}$
to the beam energy $E_b=\sqrt{s}/2$
with $s=(p_{e^-}+p_{e^+})^2$ being the center-of-mass energy squared
of the initial $e^+ e^-$ pair.
Note that $y_1y_2=\hat s/s\equiv\tau$ 
and $y_i$ varies between 0 and $x/(1+x)$ with
the machine parameter $x$ defined by
\begin{equation}
x=\frac{4E_b \omega_0}{M_e^2} = 15.3
\left(\frac{E_b}{\rm TeV}\right) \left(\frac{\omega_0}{\rm eV}\right)\,.
\end{equation}
The differential Compton cross sections are given by~\cite{Ginzburg:1982yr,Kuhn:1992fx}
\begin{equation}
\frac{{\rm d}\sigma_c^{\gamma_1}}{{\rm d}y_1} \ = \
\frac{2\pi\alpha^2}{xM_e^2}\left[f_0(y_1)+\lambda_1\,f_2(y_1)
\right]\,; \ \ \
\frac{{\rm d}\sigma_c^{\gamma_2}}{{\rm d}{y_2}} \ = \
\frac{2\pi\alpha^2}{xM_e^2}\left[\tilde f_0(y_2)+\lambda_2\,\tilde f_2(y_2)\right]\,,
\end{equation}
where, denoting $y_1=y$ and $y_2=\tau/y_1=\tau/y$,
\begin{eqnarray}
\label{eq:f0f2}
f_0(y) &=& \frac{1}{1-y}+(1-y)-4\,r\,(1-r)-r\,x\,(2\,r-1)(2-y)\,P_e P_c\,,
\nonumber \\[1mm]
f_2(y) &=& r\,x\,\left[1+(1-y)(2\,r-1)^2\right]P_e-(2\,r-1)\left(
\frac{1}{1-y}+1-y\right)P_c\,;
\nonumber \\[1mm]
\tilde f_0(\tau/y) &=& \frac{1}{1-\tau/y}+(1-\tau/y)-4\,\tilde r\,(1-\tilde r)
-\tilde r\, x\,(2\,\tilde r-1)(2-\tau/y)\tilde P_e \tilde P_c\,,
\nonumber \\[1mm]
\tilde f_2(\tau/y) &=& \tilde r\,x\,\left[1+(1-\tau/y)(2\,\tilde r-1)^2\right]\tilde P_e
-(2\,\tilde r-1)\left( \frac{1}{1-\tau/y}+1-\tau/y\right)\tilde P_c\,,
\end{eqnarray}
with 
\begin{eqnarray}
r=\frac{y}{x(1-y)}\,, \ \ \
\tilde r=\frac{\tau/y}{x(1-\tau/y)}\,.
\end{eqnarray}
The total cross sections $\sigma_c^{\gamma_1}$ and $\sigma_c^{\gamma_2}$ 
for the two Compton scattering processes are given by 
\begin{eqnarray}
\sigma_c^{\gamma_1} &=& \sigma_c^0 + P_e P_c \sigma_c^1\,; \ \ \
\sigma_c^{\gamma_2}  =  \sigma_c^0 + \tilde P_e \tilde P_c \sigma_c^1\,,
\end{eqnarray}
where the cross sections 
$\sigma_c^0$ and $\sigma_c^1$ depend on the machine parameter $x$ as
\cite{Ginzburg:1982yr}
\begin{eqnarray}
\sigma_c^0 &=& \frac{2\pi\alpha^2}{x\,M_e^2}\left[
\left(1-\frac{4}{x}-\frac{8}{x^2}\right)\log(x+1)+\frac{1}{2}+\frac{8}{x}
-\frac{1}{2(x+1)^2} \right]\,, \nonumber \\[2mm]
\sigma_c^1 &=& \frac{2\pi\alpha^2}{x\,M_e^2}\left[
\left(1+\frac{2}{x}\right)\log(x+1)-\frac{5}{2}+\frac{1}{x+1}
-\frac{1}{2(x+1)^2} \right]\,.
\end{eqnarray}
We finally obtain the following $\gamma\gamma$ luminosity of the 
two colliding photons with the helicity combination of 
$(\lambda_1,\lambda_2)$
\cite{Ginzburg:1981vm,Ginzburg:1982yr,Kuhn:1992fx,Lee:2007gz}
\footnote{The factor $4$ in the denominator of Eq.~(10) of 
Ref.~\cite{Lee:2007gz} should read $1$.} 
\begin{equation}
\label{eq:Laa}
\frac{1}{{\cal L}_{ee}}\,
\frac{{\rm d}{\cal L}^{\lambda_1\lambda_2}_{\gamma\gamma}}{{\rm d}\tau}=
\bigg(
\langle 0 0 \rangle_\tau +
\lambda_1\langle 2 0 \rangle_\tau +
\lambda_2\langle 0 2 \rangle_\tau +
\lambda_1\lambda_2\langle 2 2 \rangle_\tau \bigg)
\left(\frac{2\pi\alpha^2}{\sigma_c^{\gamma_1}\,x\,M_e^2}\right)
\left(\frac{2\pi\alpha^2}{\sigma_c^{\gamma_2}\,x\,M_e^2}\right)\,,
\end{equation}
where a convolution integral is given by
\begin{eqnarray}
\langle i j \rangle_\tau \equiv
\int_{\tau/y_{\rm max}}^{y_{\rm max}} \frac{{\rm d}y}{y}f_i(y)\tilde f_j(\tau/y)\,,
\end{eqnarray}
with $i, j = 0, 2$ and $y_{\rm max} = x/(1 + x)$.

For the unpolarized case $P_e=P_c=\tilde P_e=\tilde P_c=0$, 
only the correlation function $\langle 0 0 \rangle_\tau$ is nonvanishing, resulting
in the following helicity-independent unpolarized luminosity:
\begin{eqnarray}
\label{eq:Lunp}
\left.
\frac{1}{{\cal L}_{ee}}\,
\frac{{\rm d}{\cal L}^{\lambda_1\lambda_2}_{\gamma\gamma}}{{\rm d}\tau}
\right|_{P_e=P_c=\tilde P_e=\tilde P_c=0}=
\langle 0 0 \rangle^{\rm unp}_\tau
\left(\frac{2\pi\alpha^2}{\sigma_c^{\gamma_1}\,x\,M_e^2}\right)
\left(\frac{2\pi\alpha^2}{\sigma_c^{\gamma_2}\,x\,M_e^2}\right) \equiv
\frac{1}{{\cal L}_{ee}}\,
\frac{{\rm d}{\cal L}^{\rm unp}_{\gamma\gamma}}{{\rm d}\tau}\,,
\end{eqnarray}
where $\langle 0 0 \rangle^{\rm unp}_\tau$ represents the polarization correlation function $\langle 0 0 \rangle_\tau$ evaluated in the unpolarized case, where
$P_eP_c=\tilde P_e\tilde P_c=0$.
The differential unpolarized cross section is then found to be:
\begin{eqnarray}
\label{eq:cx_unpol_0}
\frac{{\rm d}^2\sigma_0}{{\rm d}\tau\,{\rm d}\cos\Theta} =
\frac{4}{{\cal L}_{ee}}\,
\frac{{\rm d}{\cal L}^{\rm unp}_{\gamma\gamma}}{{\rm d}\tau}\,
\frac{{\rm d}\hat\sigma_0}{{\rm d}\cos\Theta} =
\frac{\beta\,N_c}{32\pi\hat s}\,\left|{\cal A}_C\right|^2\,
\left[
\frac{4}{{\cal L}_{ee}}\,
\frac{{\rm d}{\cal L}^{\rm unp}_{\gamma\gamma}}{{\rm d}\tau}\,
\left(\widehat C_1 + \widehat C_3\right)\right]\,,
\end{eqnarray}
with the color factor $N_c=3$.
The factor 4 appears because
$\frac{1}{{\cal L}_{ee}}\,
\frac{{\rm d}{\cal L}^{\rm unp}_{\gamma\gamma}}{{\rm d}\tau}$ 
is the average of 
$\frac{1}{{\cal L}_{ee}}\,
\frac{{\rm d}{\cal L}^{\lambda_1\lambda_2}_{\gamma\gamma}}{{\rm d}\tau}$
when $P_e=P_c=\tilde P_e=\tilde P_c=0$.

To obtain the spin density matrix taking account of the helicity-dependent
$\gamma\gamma$ luminosity~\eqref{eq:Laa},
we first note that one can decompose each coefficient
$\widehat C_i$ into a sum over photon helicities as
\begin{eqnarray}
\label{eq:C_sum}
\widehat C_i = \sum_{\lambda_1,\lambda_2=\pm}
c_i^{\lambda_1\lambda_2}\,,
\end{eqnarray}
and introduce the luminosity-weighted polarization
coefficients $\widehat C_i^w$, which are defined as
\footnote{Note that,
in the unpolarized case with $P_e=P_c=\tilde P_e=\tilde P_c=0$, 
$w^{\lambda_1\lambda_2}=1/4$ since
$L^{\lambda_1\lambda_2}$ becomes independent of
$\lambda_1$ and $\lambda_2$,
see Eq.~\eqref{eq:Lunp}.
}
\begin{eqnarray}
\label{eq:C_weighted}
\widehat C_i^w
\ \equiv \ {4}\sum_{\lambda_1,\lambda_2=\pm}
w^{\lambda_1\lambda_2}\,c_i^{\lambda_1\lambda_2} \,,
\end{eqnarray}
where the weight functions are defined by
\begin{eqnarray}
\label{eq:weight_function}
w^{\lambda_1\lambda_2}\equiv \frac{L^{\lambda_1\lambda_2}}
{\sum_{\lambda_1',\lambda_2'=\pm} \ L^{\lambda_1'\lambda_2'}}\,,
\end{eqnarray}
with the luminosity functions
\begin{eqnarray}
\label{eq:w_L}
L^{\lambda_1\lambda_2}\equiv  \frac{1}{{\cal L}_{ee}}
\frac{{\rm d}{\cal L}^{\lambda_1\lambda_2}_{\gamma\gamma}}{{\rm d}\sqrt\tau}\,.
\end{eqnarray}
Explicitly, for example,
\begin{eqnarray}
\widehat C_1^w & = & \sum_{\lambda_1,\lambda_2=\pm}
\ w^{\lambda_1\lambda_2}\, 
\left[|\langle ++;\lambda_1\lambda_2\rangle|^2
+ |\langle --;\lambda_1\lambda_2\rangle|^2\right]\,,
\\ \nonumber
\widehat C_{13}^w &=&-2\,\real\sum_{\lambda_1,\lambda_2=\pm}
\ w^{\lambda_1\lambda_2}\, 
\big[\langle ++;\lambda_1\lambda_2\rangle
      \langle --;\lambda_1\lambda_2\rangle^*\big]\,.
\end{eqnarray} 

The average of the helicity-dependent $\gamma\gamma$ luminosities
is determined solely by $\langle 0 0 \rangle_\tau$, as can be checked
from Eq.~\eqref{eq:Laa}. This correlation is, in turn, a function depending
only on the products $P_eP_c$ and $\tilde P_e\tilde P_c$,
as evident in Eq.~\eqref{eq:f0f2}.
When $P_e=P_c=\tilde P_e=\tilde P_c=0$, the average reduces to
the unpolarized luminosity given by Eq.~\eqref{eq:Lunp}:
\begin{eqnarray}
\frac{1}{4}\,\sum_{\lambda_1,\lambda_2=\pm} \ L^{\lambda_1\lambda_2} =
\frac{1}{4}\,\sum_{\lambda_1,\lambda_2=\pm}
\frac{1}{{\cal L}_{ee}}
\frac{{\rm d}{\cal L}^{\lambda_1\lambda_2}_{\gamma\gamma}}{{\rm d}\sqrt\tau}
\ \  \xrightarrow{\makebox[2.1cm]{\tiny $P_eP_c=\tilde P_e\tilde P_c=0$}} \ \
\frac{1}{{\cal L}_{ee}}\,
\frac{{\rm d}{\cal L}^{\rm unp}_{\gamma\gamma}}{{\rm d}\sqrt\tau}\,.
\end{eqnarray}
%

% Collecting all the above, we arrive at the conclusion
% that the explicit form for the luminosity-weighted spin density
% matrix is still given by Eq.~\eqref{eq:SDM_hat}.
% The effect of the luminosity weighting is simply
% to replace the unweighted coefficients $\widehat C_i$ with their
% luminosity-weighted counterparts $\widehat C_i^w$
% in the expressions for the polarization vectors, spin correlations,
% and normalization factor as defined
% in Eqs.~\eqref{eq:B_hat}, \eqref{eq:C_hat}, and \eqref{eq:A_hat}.
Collecting the above results, the luminosity-weighted spin density matrix
retains the form of Eq.~\eqref{eq:SDM_hat}, with the sole effect of
luminosity weighting being the replacement of each unweighted coefficient
$\widehat{C}_i$ by its luminosity-weighted counterpart $\widehat{C}_i^w$
in the polarization vectors, spin correlations, and normalization factor
defined in Eqs.~\eqref{eq:B_hat}, \eqref{eq:C_hat}, and \eqref{eq:A_hat}.

\begin{figure}[t!]
%\begin{figure}[h!]
%\begin{figure}[b!]
%\vspace{-0.5cm}
\begin{center}
\includegraphics[width=7.5cm,height=7.5cm]{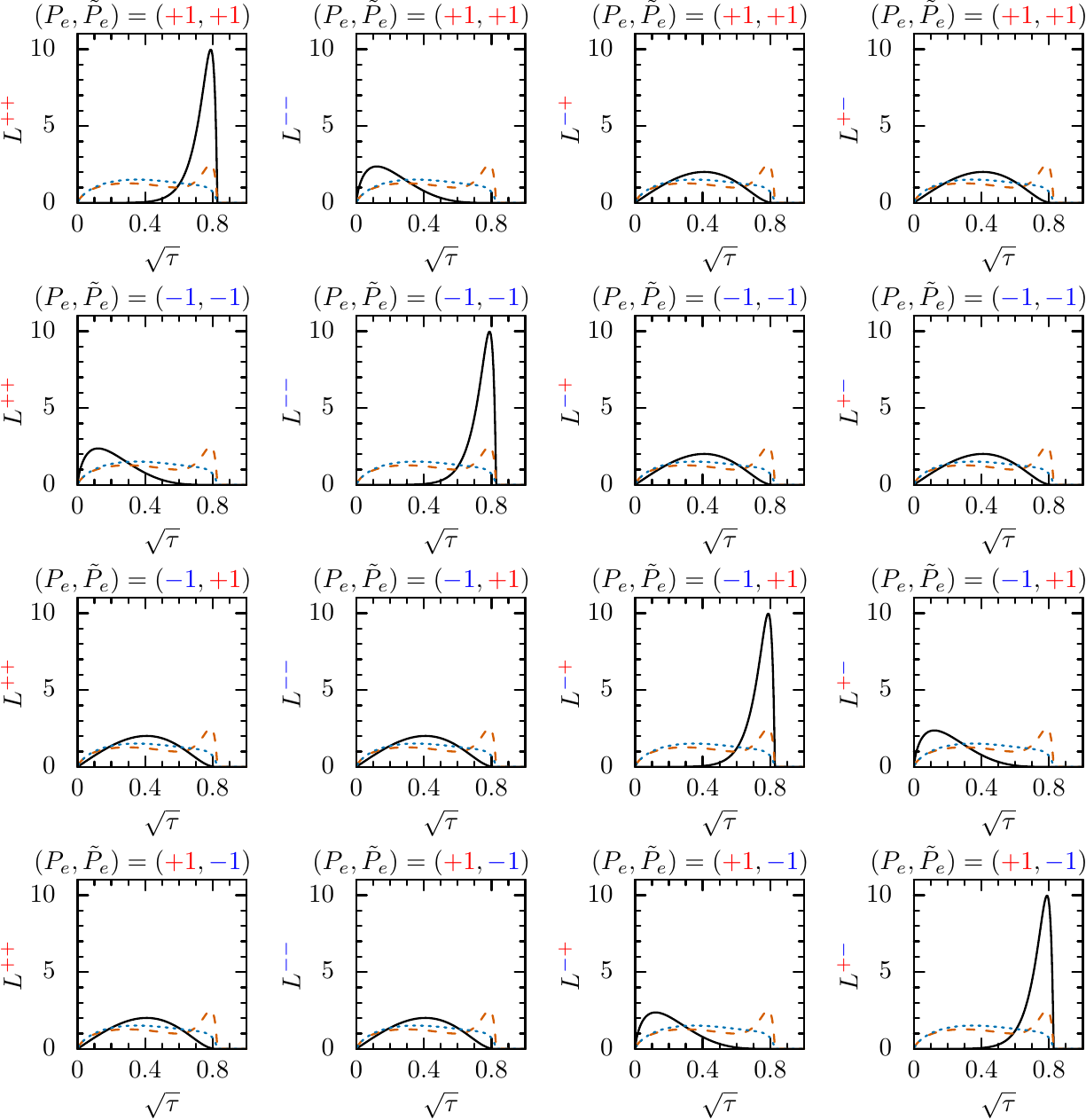}
\includegraphics[width=7.0cm,height=7.0cm]{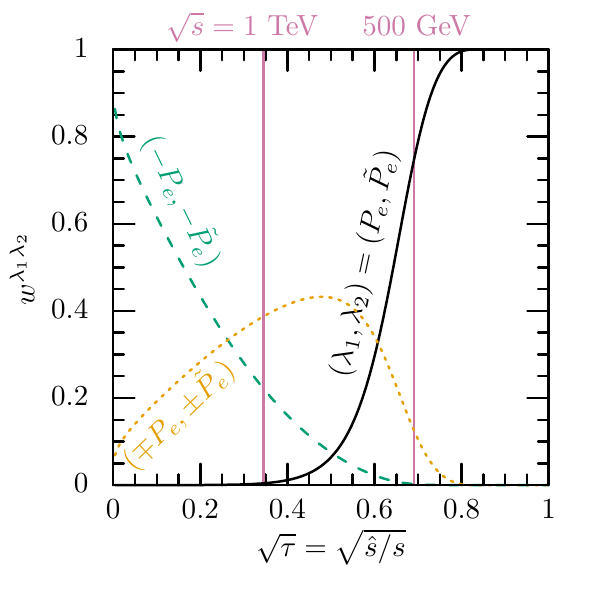}
\end{center}
\vspace{-0.5cm}
\caption{\it
(Left)
The helicity-dependent luminosities $L^{\lambda_1\lambda_2}=\frac{1}{{\cal L}_{ee}}
\frac{{\rm d}{\cal L}^{\lambda_1\lambda_2}_{\gamma\gamma}}{{\rm d}\sqrt\tau}$
taking $P_eP_c=\tilde P_e\tilde P_c=-1$ and $x=4.8$.
From left to right, $(\lambda_1,\lambda_2) =(+,+)$, $(-,-)$, $(-,+)$, and $(+,-)$
while, from top to bottom,
$(P_e=-P_c,\tilde P_e=-\tilde P_c)=(+,+)$, $(-,-)$, $(-,+)$, and $(+,-)$.
In each frame, we also show the average luminosity
$\frac{1}{4}\sum_{\lambda_1,\lambda_2=\pm}L^{\lambda_1\lambda_2}$
and the unpolarized one  $\frac{1}{{\cal L}_{ee}}
\frac{{\rm d}{\cal L}^{\rm unp}_{\gamma\gamma}}{{\rm d}\sqrt\tau}$
in dashed red and blue lines, respectively, for comparison.
(Right)
The weight functions $w^{\lambda_1\lambda_2}$ for
$(\lambda_1,\lambda_2)=(P_e,\tilde{P}_e)$ (black solid),
$(\lambda_1,\lambda_2)=(-P_e,-\tilde{P}_e)$ (green dashed), and
$(\lambda_1,\lambda_2)=(\mp P_e,\pm\tilde{P}_e)$ (orange dotted).
The vertical lines at $\sqrt{\tau}=0.345$ and $0.69$ mark the $2M_t$ threshold
for $\sqrt{s}=1~\mathrm{TeV}$ and $500~\mathrm{GeV}$, respectively.
}
\label{fig:Lum_polarized}
\end{figure}
The helicity-dependent luminosities $L^{\lambda_1\lambda_2}$ defined
in Eq.~\eqref{eq:w_L} are plotted in the left panel of Fig.~\ref{fig:Lum_polarized}
for the specific polarization configuration $P_eP_c = \tilde{P}_e\tilde{P}_c = -1$,
with the machine parameter set to $x = 4.8$.
The frames of the $4\times 4$ grid are arranged with the colliding
photon polarizations $(\lambda_1, \lambda_2) = (+,+), (-,-), (-,+), (+,-)$
displayed from left to right, and the initial electron and positron polarizations
$(P_e, \tilde{P}_e) = (+,+), (-,-), (-,+), (+,-)$
displayed from top to bottom.%
\footnote{It is understood that, for example,
$(P_e, \tilde{P}_e) = (+,+)=(+1,+1)$ under the condition $P_eP_c = \tilde{P}_e\tilde{P}_c = -1$.}
The initial laser beam polarizations, $P_c$ and $\tilde{P}_c$, are fixed
by the condition $P_eP_c = \tilde{P}_e \tilde{P}_c = -1$.
For comparison, each frame also displays the helicity-independent average luminosity,
$\sum_{\lambda_1,\lambda_2=\pm} L^{\lambda_1\lambda_2}/4$,
and the unpolarized luminosity,
$\frac{1}{{\cal L}_{ee}} \frac{{\rm d}{\cal L}^{\rm unp}_{\gamma\gamma}}{
{\rm d}\sqrt\tau}$.

We observe that the luminosity functions exhibit pronounced peaks near
$\sqrt{\tau} \simeq 0.8$ when the photon helicities are
$(\lambda_1, \lambda_2) = (P_e, \tilde{P}_e)$. This behavior clearly demonstrates
that a PLC can effectively control the polarization states
of colliding photons. Under the condition
$P_e P_c = \tilde{P}_e \tilde{P}_c = -1$, we find that more than 70\% of
the colliding photons carry the helicities of the initial electron and positron
beams within the kinematic region $0.68 \lesssim \sqrt{\tau} \lesssim 0.82$
(see the right panel of Fig.~\ref{fig:Lum_polarized}).\footnote{More precisely,
for $\lambda_1 = P_e$ and $\lambda_2 = \tilde{P}_e$, the weight function
satisfies $w^{\lambda_1\lambda_2} > 0.7$ (0.9) in the region
$0.68$ (0.73) $\lesssim \sqrt{\tau} \lesssim 0.82$.}
For the center-of-mass energy of $\sqrt{s} = 500$ (1000) GeV, this
interval corresponds to invariant masses of
$340$ (680) $\lesssim \sqrt{\hat{s}}/\text{GeV} \lesssim 410$ (820), spanning
energies near (well above) the $t \bar t$ threshold $2M_t$.
% It is worth noting that the weight function with photon
% helicities $(\lambda_1, \lambda_2) = (-P_e, +\tilde{P}_e)$ yields identical
% results to that with $(\lambda_1, \lambda_2) = (+P_e, -\tilde{P}_e)$,
% reflecting the inherent symmetry of the system.
The weight functions for
$(\lambda_1,\lambda_2)=(-P_e,+\tilde{P}_e)$ and
$(\lambda_1,\lambda_2)=(+P_e,-\tilde{P}_e)$ are identical under
$\tilde{P}_e\leftrightarrow-\tilde{P}_e$, $P_e\leftrightarrow-P_e$.

%------------------------------------------------------------------
\section{Numerical analysis}
\label{sec:Numerical}
Given the spin density matrix in Eq.~\eqref{eq:rho_0}, the four criteria for
quantum entanglement and violation of the Bell inequality, as defined in
Eqs.~\eqref{eq:PH}, \eqref{eq:Concurrence}, \eqref{eq:CHSH}, and \eqref{eq:D},
can be readily evaluated.
For our numerical analysis of the unpolarized case
($P_e=P_c=\tilde P_e=\tilde P_c=0$), we utilize the spin density matrix
in Eq.~\eqref{eq:SDM_hat}, in conjunction with
Eqs.~\eqref{eq:B_hat}, \eqref{eq:C_hat},
and \eqref{eq:A_hat}. These expressions are formulated using
the 16 polarization coefficients $\widehat C_i$ detailed in
Appendix~\ref{sec:16Cs} in terms of the reduced amplitudes defined in Eq.~\eqref{eq:HelAmp}. The LO reduced amplitudes
for the $\gamma\gamma\to t\bar t$ process are provided in
Table~\ref{tab:aatt_SM}.
For the analysis of polarized colliding photons, we introduce
the luminosity-weighted polarization coefficients $\widehat C_i^w$,
as defined in Eq.~\eqref{eq:C_weighted}. The luminosity-weighted spin
density matrix retains the form of Eq.~\eqref{eq:SDM_hat}, but
with $\widehat C_i$ replaced by $\widehat C_i^w$ in
Eqs.~\eqref{eq:B_hat}, \eqref{eq:C_hat}, and \eqref{eq:A_hat}.

This section presents a numerical analysis of the impact of colliding-photon
polarization on the observability of quantum entanglement and Bell inequality
violation in top-quark pair production at a PLC.
After studying the unpolarized case ($P_e=P_c=\tilde P_e=\tilde P_c=0$) for
comparison, we first examine scenarios with perfectly polarized photons,
where only one of the four weight functions $w^{\lambda_1\lambda_2}$
is nonzero (equal to 1), while the others vanish.
Subsequently, to address realistic cases with imperfectly polarized
colliding photons containing mixed polarization states, we perform
detailed numerical studies for two specific cases:
$\sqrt{s} = 500$ GeV with $P_e=-P_c=\tilde P_e=-\tilde P_c=+1$,
and $\sqrt{s} = 1$ TeV with $P_e=-P_c=-\tilde P_e=+\tilde P_c=+1$.

\subsection{Unpolarized colliding photons}
\begin{figure}[t!]
%\begin{figure}[h!]
%\begin{figure}[b!]
%\vspace{-0.5cm}
\begin{center}
\includegraphics[width=7.0cm]{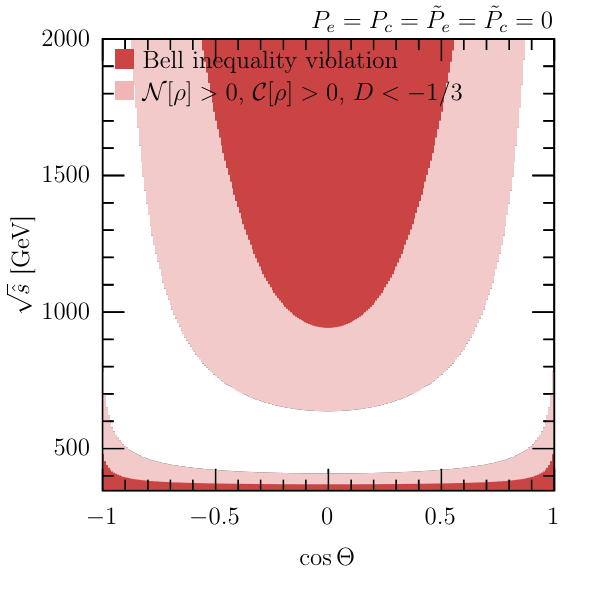}
\includegraphics[width=7.0cm]{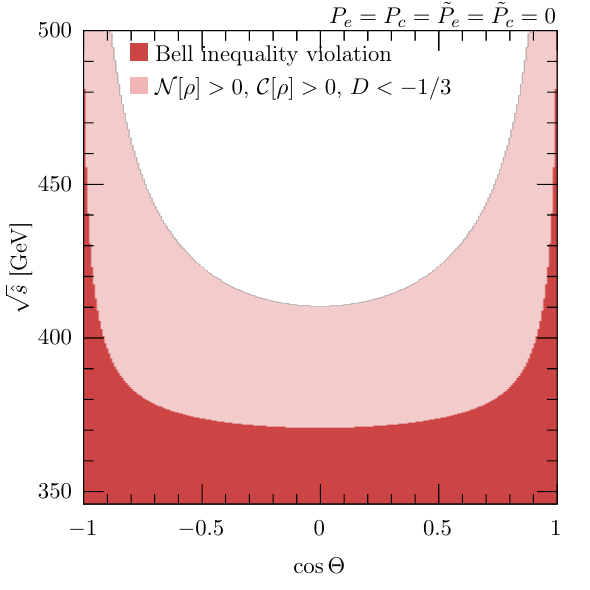}
\end{center}
\vspace{-0.5cm}
\caption{\it
{\bf Quantum entanglement in the unpolarized case}:
(Left)
Quantum entanglement
for the unpolarized case $P_e=P_c=\tilde P_e=\tilde P_c=0$,
shown in the $(\cos\Theta,\sqrt{\hat s})$ plane.
% Shaded regions indicate entanglement, satisfying ${\cal N}[\rho]>0$,
% ${\cal C}[\rho]>0$, and $D<-1/3$. This implies that the quantum state
% of the top quark pair system is entangled.
Shaded regions satisfy $\mathcal{N}[\rho]>0$,
$\mathcal{C}[\rho]>0$, and $D<-1/3$, indicating entanglement
of the $t\bar{t}$ spin state.
% In the thickly shaded red regions, we further have
% $m_{12}>1$ indicating the violation of the Bell inequality.
The densely shaded red regions additionally satisfy $m_{12}>1$,
indicating violation of the Bell inequality.
(Right)
Magnified view of the region $2M_t < \sqrt{\hat s} < 500$~GeV.
}
\label{fig:QE_unp}
\end{figure}
Figure~\ref{fig:QE_unp} shows the entanglement region in the
$(\cos\Theta,\sqrt{\hat{s}})$ plane for the unpolarized case
($P_e=P_c=\tilde{P}_e=\tilde{P}_c=0$).
Shaded regions satisfy $\mathcal{N}[\rho]>0$, $\mathcal{C}[\rho]>0$, and $D<-1/3$;
densely shaded red regions additionally satisfy $m_{12}>1$,
indicating violation of the Bell inequality.
As expected, the criteria ${\cal N}[\rho]>0$ and ${\cal C}[\rho]>0$ identify
the same regions, and the condition $D<-1/3$ is equally effective
at detecting entanglement. The Bell inequality is violated near the $2M_t$
threshold and in the region where $\sqrt{\hat s}\gsim 1$ TeV and
$|\cos\Theta| \lsim 0.5$.

\subsection{Perfectly polarized colliding photons with \boldmath{$w^{\pm\pm}=1$} }

To analyze the scenario with perfectly polarized colliding photons
of the same helicity, we first consider the case where $w^{++}=1$ and, accordingly,
$w^{--}=w^{+-}=w^{-+}=0$. In this configuration with $\lambda_1=\lambda_2=+$,
the luminosity-weighted polarization coefficients $\widehat C_i^w$ receive
contributions only from the two reduced amplitudes with
$\sigma\bar\sigma=\pm\pm$ (see Table~\ref{tab:aatt_SM}):
\begin{equation}
\langle ++;++ \rangle = \frac{2M_t}{\sqrt{\hat s}} (1+\beta)\,, \ \ \
\langle --;++ \rangle = \frac{2M_t}{\sqrt{\hat s}} (1-\beta)\,.
\end{equation}
Using the explicit expressions for the polarization coefficients
in Appendix~\ref{sec:16Cs}, we find that only three of the 16
(luminosity-weighted) polarization coefficients are nonzero:
\begin{eqnarray}
\widehat C^w_1 = {8}\frac{M_t^2}{\hat s} (1+\beta^2)\,, \ \ \
\widehat C^w_2 = {16}\frac{M_t^2}{\hat s} \beta\,, \ \ \
\widehat C^w_{13} = -{8}\frac{M_t^2}{\hat s} (1-\beta^2)\,.
\end{eqnarray}
Consequently, based on Eqs.~\eqref{eq:SDM_hat}, \eqref{eq:B_hat}, \eqref{eq:C_hat},
and \eqref{eq:A_hat}, we find that only the following polarization
vectors and spin correlations are nonvanishing:
\begin{equation}
B^+_k = - B_k^- = \frac{\widehat C^w_2}{\widehat C^w_1} =
\frac{2\beta}{1+\beta^2}\,, \ \ \
C_{nn} = C_{rr} = \frac{\widehat C^w_{13}}{\widehat C^w_1} =
-\frac{1-\beta^2}{1+\beta^2} < 0 \,, \ \ \
C_{kk} = -\frac{\widehat C^w_{1}}{\widehat C^w_1} =  -1 < 0 \,.
\end{equation}
Substituting these polarization vectors and spin correlations
into Eq.~\eqref{eq:rho_gen}\footnote{We identify that
$1=\hat n$, $2=\hat r$, and $3=\hat k$.},
we obtain the following spin density matrices $\rho$ and
$\rho^{T_B}$:
\begin{eqnarray}
\hskip -0.3cm
\rho &=& \frac{1}{2(1+\beta^2)}\left(
\begin{array}{cc|cc}
0&0&0&0 \\ 0 & (1+\beta)^2 & -(1-\beta^2) & 0 \\ \hline
0 & -(1-\beta^2) & (1-\beta)^2 & 0 \\ 0 & 0& 0& 0
\end{array}\right)\,; \,
\nonumber \\
\rho^{T_B} &=& \frac{1}{2(1+\beta^2)}\left(
\begin{array}{cc|cc}
0&0&0&-(1-\beta^2) \\ 0 & (1+\beta)^2 & 0 & 0 \\ \hline
0 & 0 & (1-\beta)^2 & 0 \\ -(1-\beta^2) & 0& 0& 0
\end{array}\right)\,.
\end{eqnarray}
Note that $\rho^{T_B}$ is obtained by transposing the four $2\times 2$
blocks of $\rho$.
%
% We observe that,
In the limit $\beta\to 0$, $B_i^+=B_j^-=0$ and
$C_{ij}=-\delta_{ij}$; consequently, the spin density matrix $\rho$
represents a pure singlet state with $J^P=0^-$,\footnote{For a CP-even scalar
with $J^P=0^+$, $B_i^+=B_j^-=0$, $C_{nn}=C_{rr}=+1$, and $C_{kk}=-1$.}
satisfying $\rho^2=\rho$, which indicates perfectly anti-correlated spins
between the two spin-$1/2$ particles~\cite{Barr:2024djo}.
The four eigenvalues of $\rho^{T_B}$ are
\begin{equation}
\pm\,\frac{1-\beta^2}{2(1+\beta^2)}\,, \ \
\frac{(1-\beta)^2}{2(1+\beta^2)}\,, \ \
\frac{(1+\beta)^2}{2(1+\beta^2)}\,.
\end{equation}
We observe that $\rho^{T_B}$ has three positive eigenvalues. However, the fourth
eigenvalue is always negative and approaches 0 as $\beta \to 1$, yielding the
negativity
\begin{equation}
{\cal N}[\rho] \ = \ \frac{1-\beta^2}{2(1+\beta^2)} > 0\,.
\end{equation}
This result guarantees that quantum entanglement is present
throughout the entire $(\cos\Theta,\sqrt{\hat{s}})$ plane
for $w^{++}=1$, as a consequence of the Peres-Horodecki
criterion~\eqref{eq:PH}.

The $4\times 4$ auxiliary matrix $R$, which is closely linked to
concurrence ${\cal C}[\rho]$ of the system, is given by
\begin{equation}
R \ = \
\rho\,\left(\sigma_2\otimes\sigma_2\right)\,
\rho^*\,\left(\sigma_2\otimes\sigma_2\right)
\ = \ \frac{1}{2(1+\beta^2)^2}\left(
\begin{array}{cc|cc}
0&0&0&0 \\ 0 & (1-\beta^2)^2 & -(1+\beta)^2(1-\beta^2) & 0 \\ \hline
0 & -(1-\beta)^2(1-\beta^2) & (1-\beta^2)^2 & 0 \\ 0 & 0& 0& 0
\end{array}\right)\,,
\end{equation}
when $w^{++}=1$. Its four eigenvalues are
\begin{equation}
r_1^2 \ = \ \frac{(1-\beta^2)^2}{(1+\beta^2)^2} \,, \ \ \
r_2^2 =r_3^2 = r_4^2 \ = \ 0\,.
\end{equation}
As a consequence of the analytic results derived above, we can obtain
the following expression for concurrence ${\cal C}[\rho]$ of the quantum state:
\begin{equation}
{\cal C}[\rho] \ = \ r_1 \ = \ \frac{1-\beta^2}{1+\beta^2}
\ = \ 2\,{\cal N}[\rho] \ > \ 0 \,.
\end{equation}
This result further confirms quantum entanglement throughout
the $(\cos\Theta\,, \sqrt{\hat s})$ plane for $w^{++}=1$,
consistent with Eq.~\eqref{eq:Concurrence}.

%
%\begin{figure}[t!]
\begin{figure}[h!]
%\begin{figure}[b!]
%\vspace{-0.5cm}
\begin{center}
\includegraphics[width=9.9cm,height=9.9cm]{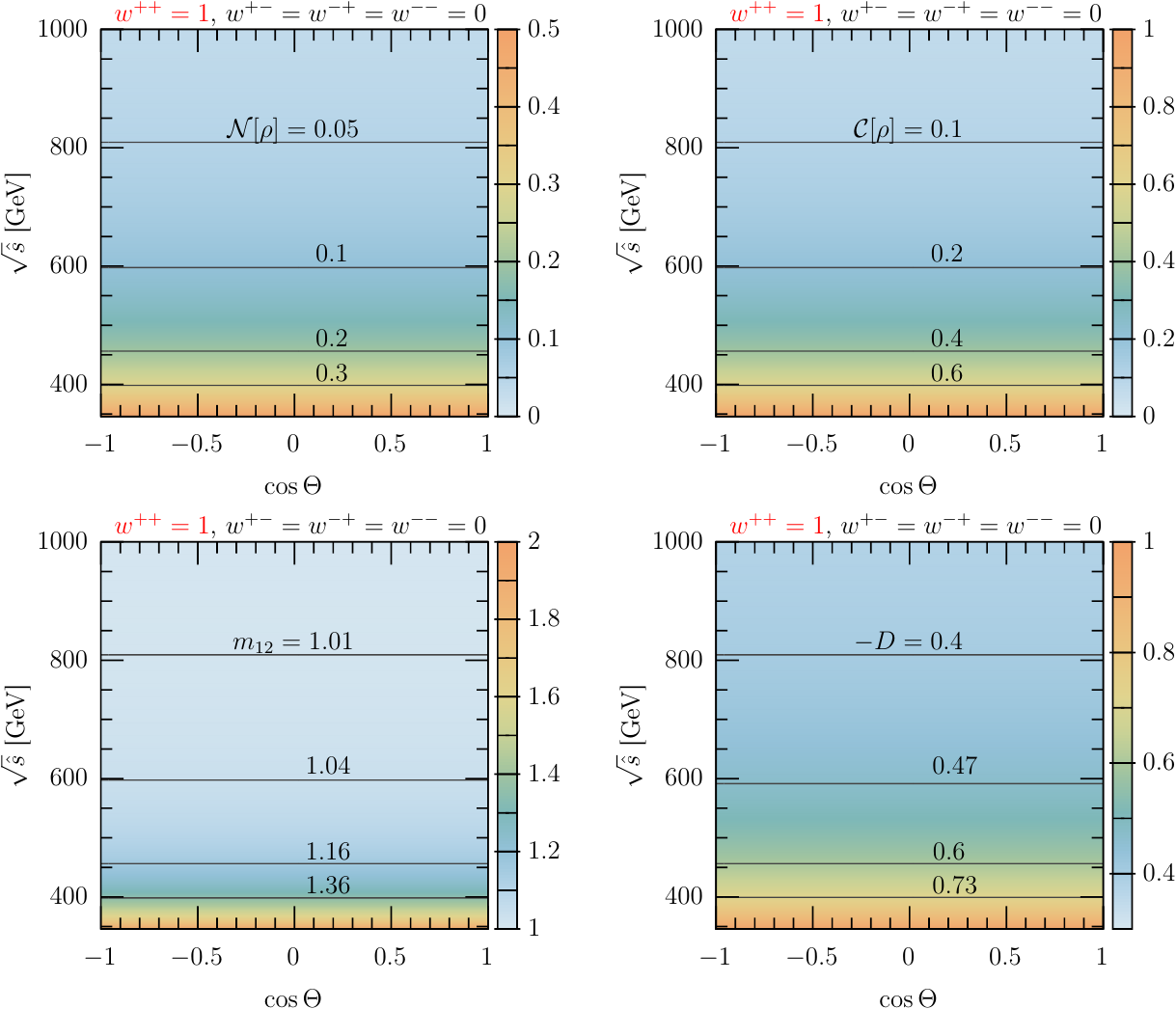}
%\hskip 0.3cm
\includegraphics[width=4.5cm,height=9.9cm]{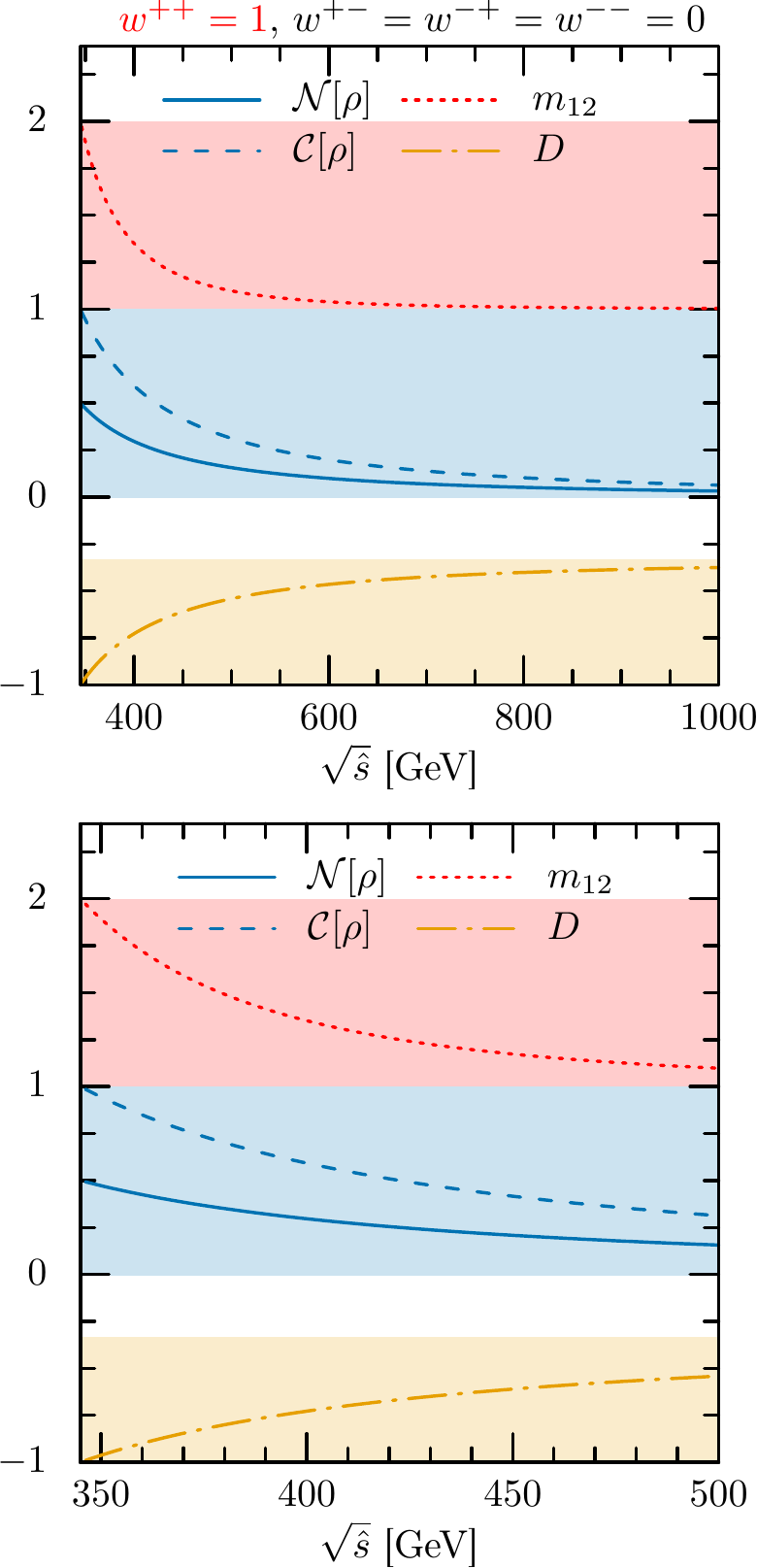}
\end{center}
%\vspace{-0.5cm}
\caption{\it
{\bf Quantum entanglement with \boldmath{$w^{\pm\pm}=1$}}:
(Left and Middle) 
Contour regions in the $(\cos\Theta,\sqrt{\hat s})$ plane for
negativity ${\cal N}[\rho]$ (upper-left),
concurrence ${\cal C}[\rho]$ (upper-right),
the Bell nonlocality parameter $m_{12}$, and
the entanglement marker $D$ (lower-right).
(Right)
Bell nonlocality parameter $m_{12}$,
concurrence $\mathcal{C}[\rho]$, negativity $\mathcal{N}[\rho]$,
and entanglement marker $D$ as functions of $\sqrt{\hat{s}}$,
from top to bottom.
Shaded regions satisfy $m_{12}>1$, $\mathcal{C}[\rho]>0$,
$\mathcal{N}[\rho]>0$, and $D<-1/3$,
indicating quantum entanglement and violation of the Bell inequality.
The lower frame magnifies the region $2M_t<\sqrt{\hat s}<500$ GeV.
}
\label{fig:QE_perfectPP}
\end{figure}

Concerning the CHSH criterion, as expressed in Eq.~\eqref{eq:CHSH},
for the Bell nonlocality parameter $m_{12}$, we find that:
\begin{equation}
C \ = \ {\rm diag}\left(-\frac{1-\beta^2}{1+\beta^2},
-\frac{1-\beta^2}{1+\beta^2},-1 \right)\,,
\end{equation}
which leads to the sum of the two largest eigenvalues of the matrix $CC^T$:
\begin{equation}
m_{12} \ = \ m_1+m_2 = 1+ \frac{(1-\beta^2)^2}{(1+\beta^2)^2}
\ = \ 1+\left({\cal C}[\rho]\right)^2 \ > \ 1\,.
\end{equation}
This inequality also ensures Bell inequality violation across
the entire $(\cos\Theta, \sqrt{\hat{s}})$ plane for $w^{++}=1$.
For the entanglement marker $D$ in Eq.~\eqref{eq:D},
we have
\begin{equation}
D=\frac{\Delta_E}{3}=
\frac{1}{3}\left(C_{nn}- | C_{rr}+C_{kk} |\right)
=\frac{1}{3}{\rm Tr}[C]
=\frac{1}{3}\left(-1-2\,\frac{1-\beta^2}{1+\beta^2}\right)
=\frac{1}{3}\left(-1-2\,{\cal C}[\rho]\right)
< -\frac{1}{3}\,,
\end{equation}
because $C_{nn}-|C_{rr}+C_{kk}|={\rm Tr}[C]$ with $C_{nn,rr,kk} <0$.

The left and middle panels of Fig.~\ref{fig:QE_perfectPP} show
the contour regions in the $(\cos\Theta\,,\sqrt{\hat s})$ plane for
negativity ${\cal N}[\rho]$~\eqref{eq:Negativity} (upper left),
concurrence ${\cal C}[\rho]$~\eqref{eq:Concurrence} (upper right),
the Bell nonlocality parameter $m_{12}$~\eqref{eq:CHSH} (lower left), and
the entanglement marker $D$~\eqref{eq:D} (lower right),
for perfectly polarized colliding photons with $w^{++}=1$.%
\footnote{We plot $-D$ instead of $+D$ so that larger values of ${\cal N}[\rho]$,
${\cal C}[\rho]$, $m_{12}$, and $-D$ consistently indicate stronger entanglement.}
The upper frame of the right panel presents plots of $m_{12}$, ${\cal C}[\rho]$,
${\cal N}[\rho]$, and $D$ (arranged from top to bottom) as functions of
$\sqrt{\hat s}$. The region $2M_t<\sqrt{\hat s}<500$ GeV is magnified
in the lower frame.
We observe that all the entanglement quantifiers are essentially independent
of $\cos\Theta$. At $\sqrt{\hat s}=2M_t$, they reach values indicative of
maximal entanglement and approach the boundary between entanglement
and separability as $\sqrt{\hat s}\to\infty$.
Importantly, all four criteria for quantum entanglement and Bell inequality
violation, namely ${\cal N}[\rho]>0$, ${\cal C}[\rho]>0$, $m_{12}>1$,
and $-D>1/3$, are satisfied across the entire range of $\sqrt{\hat s}$,
although the entanglement quantifiers decrease rapidly with increasing
$\sqrt{\hat s}$.
This behavior is precisely captured by the analytical expressions
derived above, demonstrating the utility of the spin density matrix
formalism for a two-qubit system. Constructed from the amplitude level,
the formalism is particularly useful at incorporating
polarization effects at colliders.

Finally, we observe that, for $w^{--}=1$ and $w^{++}=w^{+-}=w^{-+}=0$,
the luminosity-weighted coefficients $\widehat C^w_1$ and $\widehat C^w_{13}$
are unaffected, whereas $\widehat C^w_2$ changes sign. Consequently,
the entanglement quantifiers ${\cal N}[\rho]$, ${\cal C}[\rho]$, $m_{12}$,
and $D$ remain the same, ensuring quantum entanglement throughout
the $(\cos\Theta, \sqrt{\hat s})$ plane, as in the $w^{++}=1$ case.

\subsection{Perfectly polarized colliding photons with
\boldmath{$w^{\pm\mp}=1$} }
%

%\begin{figure}[t!]
%\begin{figure}[h!]
\begin{figure}[b!]
%\vspace{-0.5cm}
\begin{center}
\includegraphics[width=9.9cm,height=9.9cm]{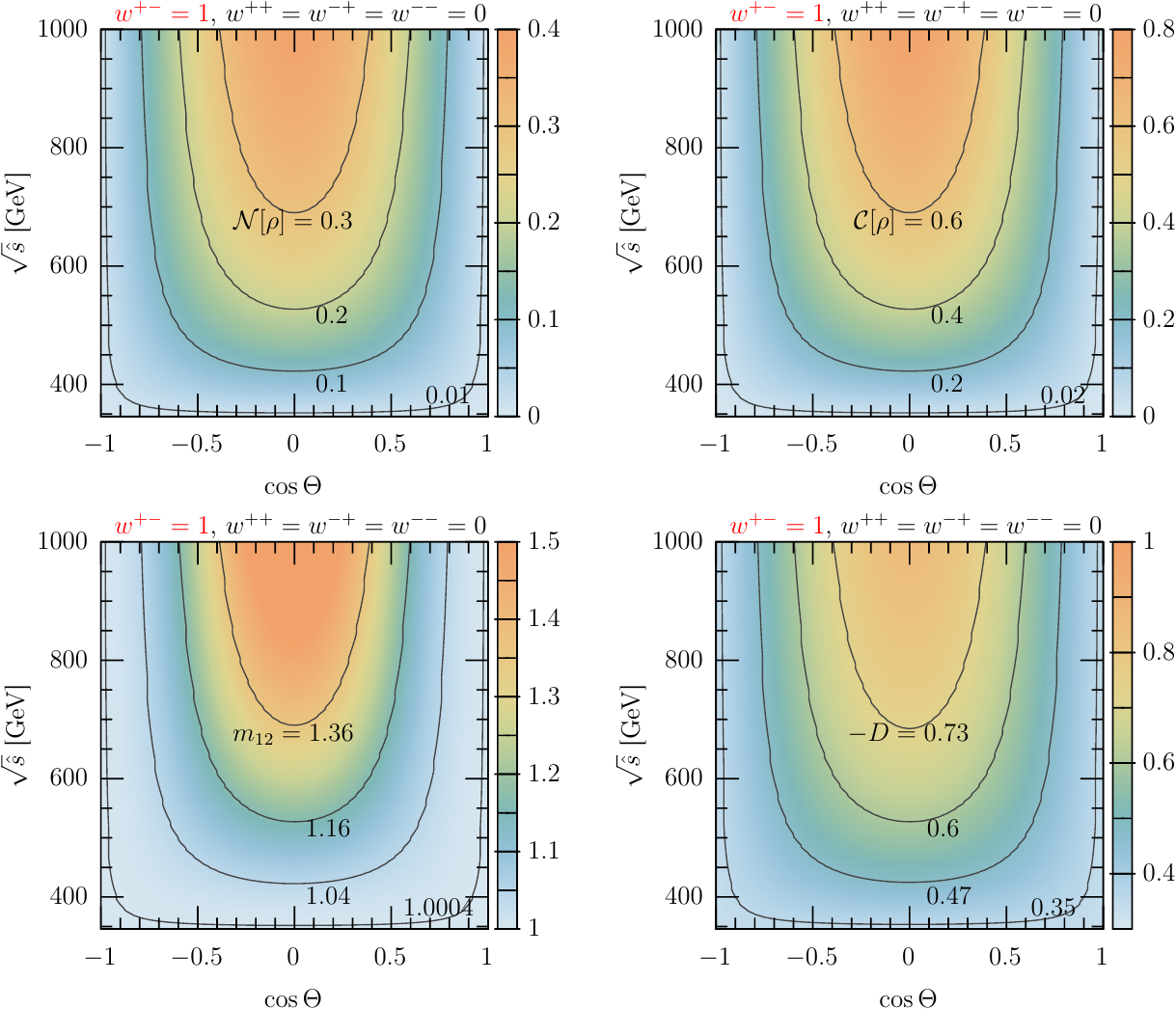}
\hskip 0.3cm
\includegraphics[width=4.5cm,height=9.9cm]{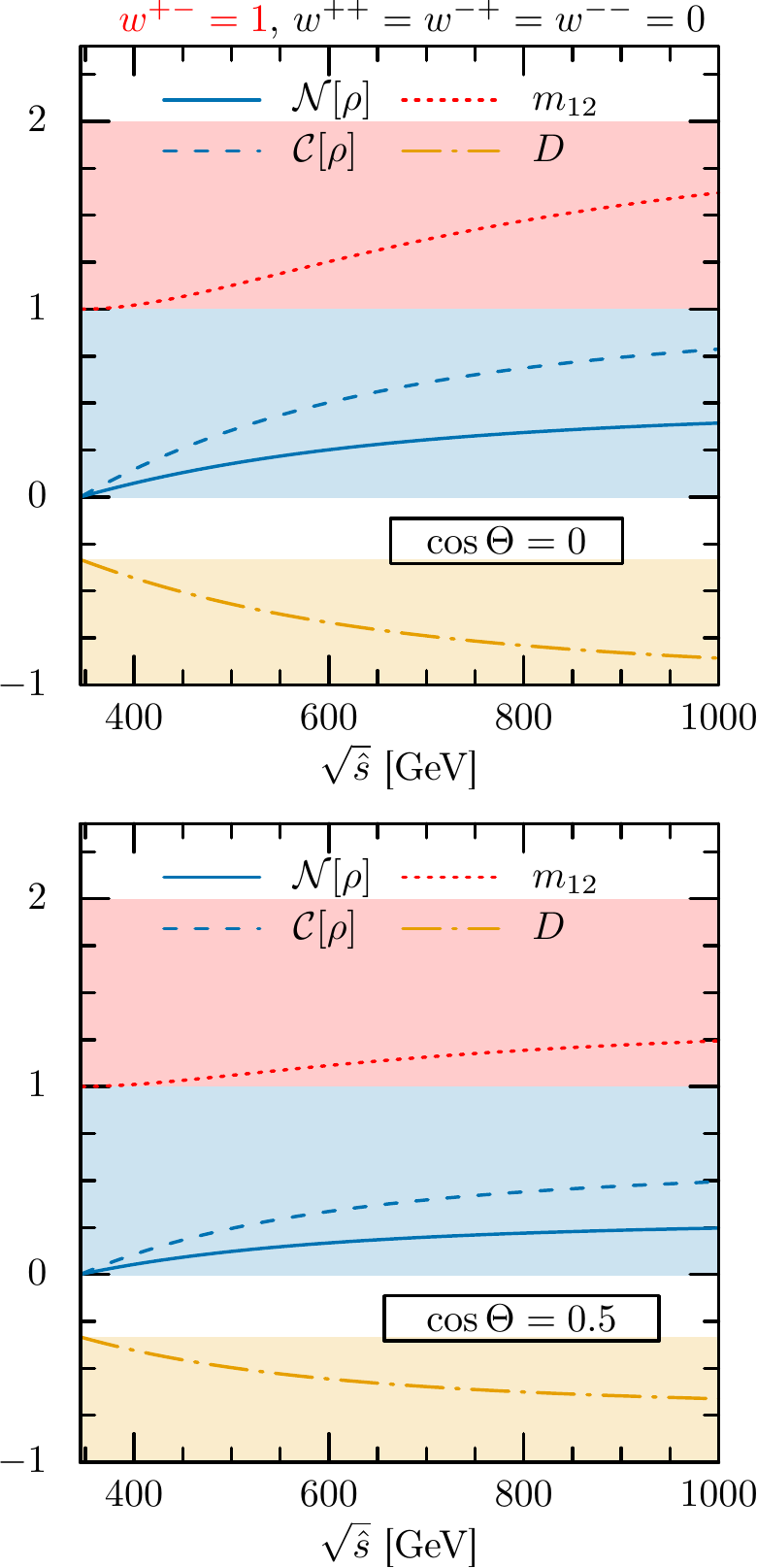}
\end{center}
\vspace{-0.5cm}
\caption{\it
{\bf Quantum entanglement with \boldmath{$w^{\pm\mp}=1$}}:
Same as in Fig.~\ref{fig:QE_perfectPP} but for
$w^{\pm\mp}=1$ and, in the right panel,
for $\cos\Theta=0$ and $0.5$.
}
\label{fig:QE_perfectPM}
\end{figure}
The situation becomes more complex when the colliding photons have
opposite helicities, with $w^{\pm\mp}=1$ and $w^{++}=w^{--}=w^{\mp\pm}=0$.
In this case, the luminosity-weighted polarization coefficients
$\widehat C_i^w$ receive contributions from all four reduced amplitudes with
$\sigma\bar\sigma = ++, --, -+$, and $+-$.
To illustrate, consider the specific case where
$w^{+-}=1$ at $\cos\Theta=0$ and $\sin\Theta=1$.
For $\lambda_1=-\lambda_2=+1$, the luminosity-weighted polarization
coefficients $\widehat C_i^w$ receive contributions from the following
reduced amplitudes (see Table~\ref{tab:aatt_SM}):
\begin{equation}
\langle ++;+- \rangle =-\frac{2M_t}{\sqrt{\hat s}}\,\beta\,, \ \
\langle --;+- \rangle = \frac{2M_t}{\sqrt{\hat s}}\,\beta\,; \ \ \
\langle -+;+- \rangle = \beta\,, \ \
\langle +-;+- \rangle =-\beta\,.
\end{equation}
Specifically, we observe that the following 5 luminosity-weighted polarization
coefficients are nonzero:\footnote{Note that if $w^{-+}=1$, $\widehat C^w_5$
changes sign, whereas the other coefficients are unaffected.}
\begin{eqnarray}
\widehat C^w_1 = {8}\frac{M_t^2}{\hat s} \beta^2\,, \ \ \
\widehat C^w_3 = {2}\,\beta^2\,, \ \ \
\widehat C^w_5 =-\frac{8M_t}{\sqrt{\hat s}}\,\beta^2\,, \ \ \
\widehat C^w_{13} = {8}\frac{M_t^2}{\hat s} \beta^2\,, \ \ \
\widehat C^w_{15} = {2}\,\beta^2\,.
\end{eqnarray}
Therefore, based on Eqs.~\eqref{eq:SDM_hat}, \eqref{eq:B_hat}, \eqref{eq:C_hat}, and
\eqref{eq:A_hat}, we observe that only the following polarization
vectors and spin correlations are nonvanishing:
\begin{equation}
B^+_r = B_r^- = \frac{4M_t\sqrt{\hat s}}{\hat s +4M_t^2}\,, \ \ \
C_{nn} = -\frac{\hat s-4 M_t^2}{\hat s+4M_t^2}<0\,, \ \ \
C_{rr} = 1>0\,, \ \ \
C_{kk} = -C_{nn}>0\,.
\end{equation}
We note that the diagonal $3\times 3$ spin correlation matrix,
given by $C={\rm diag}\left(C_{nn},C_{rr},C_{kk}\right)$, leads to
\begin{eqnarray}
\left.m_{12}\right|_{\cos\Theta=0}  &=&
1 \ + \ \frac{(\hat s-4 M_t^2)^2}{(\hat s+4M_t^2)^2} \ > \ 1\,; 
\nonumber \\
\left.D\right|_{\cos\Theta=0} &=& \frac{1}{3}\left(C_{nn}- | C_{rr}+C_{kk} |\right)
=\frac{1}{3}\left(-1-2\,\frac{\hat s-4M_t^2}{\hat s+4M_t^2}\right)
< -\frac{1}{3}\,.
\end{eqnarray}
In fact, for $w^{\pm\mp}=1$, we find that the entanglement quantifiers are
\begin{equation}
{\cal N}[\rho]=\frac{\beta^2\,\sin^2\Theta}{2(2-\beta^2\,\sin^2\Theta)}\,, \ \
{\cal C}[\rho]=2\,{\cal N}[\rho]\,, \ \
m_{12}= 1 + \big({\cal C}[\rho]\big)^2\,, \ \
D=\frac{1}{3}\big(-1-2\,{\cal C}[\rho]\big)\,.
\end{equation}
For a given value of $\sqrt{\hat s}$, the strongest entanglement occurs
at $\sin\Theta=1$, where ${\cal C}[\rho]
=2{\cal N}[\rho]=\beta^2/(2-\beta^2) = (\hat s -4 M_t^2)/(\hat s +4 M_t^2)$.
In contrast to the $w^{\pm\pm}=1$ case, 
we find that
the boundary between entanglement
and separability is located at $\sqrt{\hat s}= 2M_t$ and
all entanglement quantifiers approach their maximal
values as $\sqrt{\hat s}\to\infty$, when $w^{\pm\mp}=1$.
In this limit of $\sqrt{\hat s}\to\infty$, $B^+_r = B^-_r = 0$ and 
$C = {\rm diag}(-1,1,1)$ at $\cos\Theta=0$; 
consequently, the spin density matrix $\rho$ represents a pure triplet state.

Similar to Fig.~\ref{fig:QE_perfectPP}, but now for $w^{+-}=1$,
the left and middle panels of Fig.~\ref{fig:QE_perfectPM} display
contour regions in the $(\cos\Theta,\sqrt{\hat s})$ plane for
negativity ${\cal N}[\rho]$, concurrence ${\cal C}[\rho]$,
the Bell nonlocality parameter $m_{12}$, and the entanglement marker $D$.
The right panel shows the $\sqrt{\hat s}$ dependence of these quantities
at $\cos\Theta=0$ (upper) and $\cos\Theta=0.5$ (lower).
Note that ${\cal N}[\rho]$, ${\cal C}[\rho]$, $m_{12}$, and $-D$
increase slowly as $\sqrt{\hat s}$ increases, in contrast to the same-helicity
case, where these quantities decrease rapidly as $\sqrt{\hat s}$ moves
away from the $2M_t$ threshold.
This makes opposite-helicity photons the preferred choice for
entanglement measurement over a broad range of $\sqrt{\hat s}$,
particularly in the high-$\sqrt{\hat s}$ region, whereas
same-helicity photons remain advantageous near the $2M_t$ threshold.

\subsection{Polarized colliding photons
with \boldmath{$\sqrt s =500$} GeV and \boldmath{$P_e=\tilde P_e=+1$}}
%
%
%\begin{figure}[t!]
%\begin{figure}[h!]
\begin{figure}[b!]
%\vspace{-0.5cm}
\begin{center}
\includegraphics[width=9.9cm,height=9.9cm]{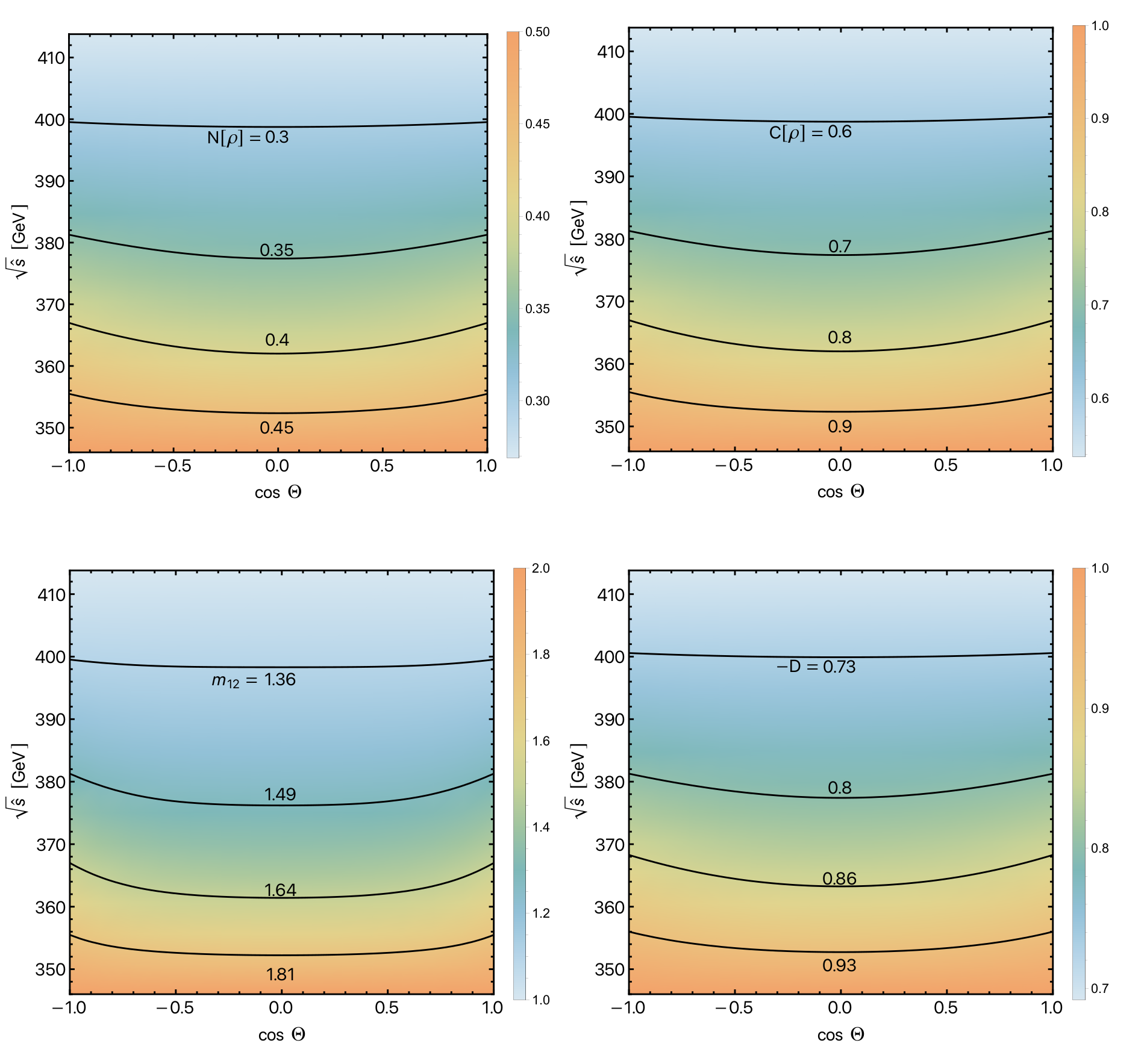}\hskip -0.5cm
\includegraphics[width=4.5cm,height=9.9cm]{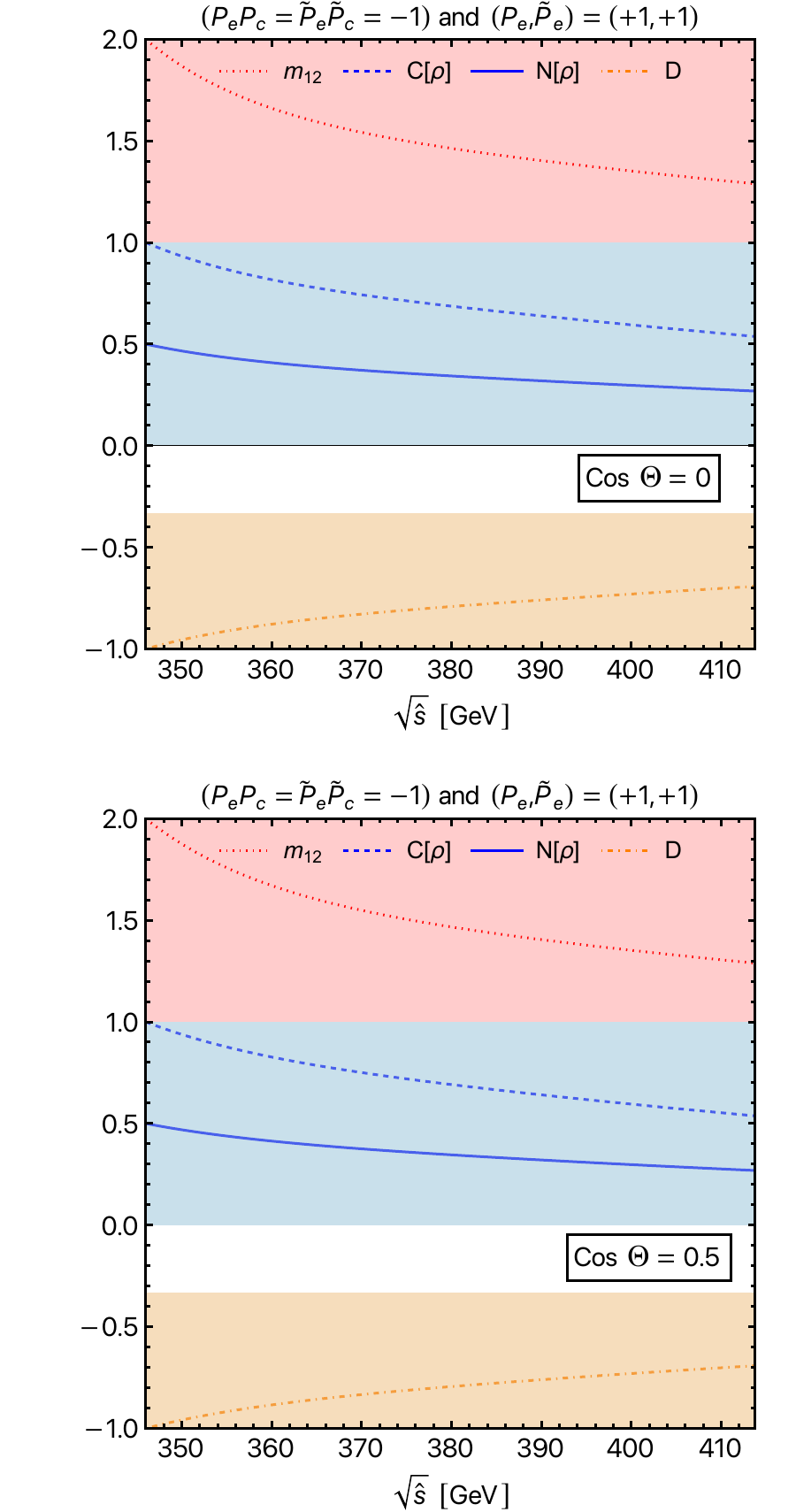}
\end{center}
\vspace{-0.5cm}
\caption{\it
{\bf Quantum entanglement with \boldmath{$\sqrt s= 500$} GeV and
\boldmath{$P_e=\tilde P_e=+1$}}:
Same as in Fig.~\ref{fig:QE_perfectPM} but for
$\sqrt s= 500$ GeV and $P_e=-P_c=\tilde P_e=-\tilde P_c=+1$.
}
\label{fig:QE_500GeV}
\end{figure}

To address the case of imperfectly polarized colliding photons
near the $2M_t$ threshold, we consider the case where
\begin{equation}
\sqrt{s} \ = \ 500 \ {\rm GeV}\,; \ \
P_e=-P_c=\tilde P_e=-\tilde P_c=+1\,.
\end{equation}
In this case, the invariant mass of the $t\bar t$ system is constrained
by $2M_t  <  \sqrt{\hat s}  \lsim  410 \ {\rm GeV}$, and the colliding
photons are polarized with $w^{++} \gsim  0.7$, $w^{+-}=w^{-+} \lsim 0.15$,
and $w^{--}  \lsim  3\times 10^{-3}$ (see the right panel of Fig.~\ref{fig:Lum_polarized}). We find that quantum entanglement is
observable throughout the entire $(\cos\Theta, \sqrt{\hat s})$ plane,
as this case can be well approximated by the perfectly polarized scenario
with $w^{++}=1$.

Figure~\ref{fig:QE_500GeV} displays the contour regions for negativity
${\cal N}[\rho]$, concurrence ${\cal C}[\rho]$, the Bell nonlocality
parameter $m_{12}$, and the entanglement marker $D$.
We recall that, for given values of $\cos\Theta$ and $\sqrt{\hat s}$,
the entanglement quantifiers are calculated using the spin density
matrix $\rho$ from Eq.~\eqref{eq:SDM_hat}, with $\widehat C_i$ replaced
by $\widehat C_i^w$ in Eqs.~\eqref{eq:B_hat}, \eqref{eq:C_hat},
and \eqref{eq:A_hat}.
The luminosity-weighted polarization coefficients $\widehat C_i^w$
are defined in Eq.~\eqref{eq:C_weighted}.
For example, the $kk$ element of the $3\times 3$ spin correlation matrix $C$ is
given by
\begin{eqnarray}
C_{kk}\left(\sqrt{\hat s}\,,\cos\Theta\right)=
\frac{\widehat{C}_{kk}}{\widehat{A}}=
\frac{-\widehat{C}_{1}^w+\widehat{C}_{3}^w}
{\widehat{C}_{1}^w+\widehat{C}_{3}^w}\,.
\end{eqnarray}
Since $w^{++}\gsim 0.7$ with $\sqrt{s}=500$ GeV,
the accessible range is confined to
$\sqrt{\hat s} \lsim 410 \ {\rm GeV}$, corresponding to
the narrow region near threshold in Fig.~\ref{fig:QE_perfectPP}.
We observe that all four entanglement quantifiers are nearly independent of
$\cos\Theta$. At $\sqrt{\hat s}=2M_t$, they attain their maximal entanglement
values: negativity ${\cal N}[\rho]=1/2$, concurrence ${\cal C}[\rho]=1$, the
Bell nonlocality parameter $m_{12}=2$, and $D=-1$.
%%
%In Fig.~\ref{fig:NCm12D_1d},
%we show negativity ${\cal N}[\rho]$,
%concurrence ${\cal C}[\rho]$,
%the Bell nonlocality parameter $m_{12}$, and
%the entanglement marker $D$
%as functions of $\sqrt{\hat s}$ (left) and $\cos\Theta$  (right)
%which have been obtained by using
%the polarization vectors and spin correlations
%with one of the kinematic variables integrated.
%%
%To be explicit, for example,
%\begin{eqnarray}
%C_{kk}\left(\sqrt{\hat s}\right)&=&\dfrac
%{\int_{-1}^{+1}
%\frac{\beta\,N_c}{32\pi\hat s}\,\left|{\cal A}_C\right|^2\,
%\left(-\widehat{C}_{1}^w+\widehat{C}_{3}^w\right)\,{\rm d}\cos\Theta}
%{\int_{-1}^{+1}
%\frac{\beta\,N_c}{32\pi\hat s}\,\left|{\cal A}_C\right|^2\,
%\left(\widehat{C}_1^w+\widehat{C}_3^w\right)\,{\rm d}\cos\Theta}\,,
%\nonumber \\[2mm]
%C_{kk}\left(\cos\Theta\right)&=&\dfrac
%{\int_{2M_t}^{\left.\sqrt{\hat s}\right|_{\rm max}}\,
%\frac{\beta\,N_c}{32\pi\hat s}\,\left|{\cal A}_C\right|^2\,
%\left(-\widehat{C}_{1}^w+\widehat{C}_{3}^w\right)\,{\rm d}\sqrt{\hat s}}
%{\int_{2M_t}^{\left.\sqrt{\hat s}\right|_{\rm max}}\,
%\frac{\beta\,N_c}{32\pi\hat s}\,\left|{\cal A}_C\right|^2\,
%\left(\widehat{C}_1^w+\widehat{C}_3^w\right)\,{\rm d}\sqrt{\hat s}}\,,
%\end{eqnarray}
%with
%\begin{eqnarray}
%{\cal A}_C =
%\frac{8\pi\alpha Q_t^2}{1-\beta^2 \cos^2\Theta} \ \ \ {\rm and} \ \ \
%\beta=\sqrt{1-\frac{4M_t^2}{\hat s}}\,.
%\end{eqnarray}

\subsection{Polarized colliding photons with \boldmath{$\sqrt s =1$} TeV and
\boldmath{$P_e=-\tilde P_e=+1$}}

To address the high-$\sqrt{\hat s}$ regime and to cover a broader range of
$\sqrt{\hat s}$, we now consider the case where
\begin{equation}
\sqrt{s} \ = \ 1 \ {\rm TeV}\,; \ \
P_e=-P_c=-\tilde P_e=+\tilde P_c=+1\,.
\end{equation}
In this scenario, the invariant mass of the $t\bar t$ system is constrained
to $2M_t  <  \sqrt{\hat s}  \lsim  820 \ {\rm GeV}$.
In the region $2M_t < \sqrt{\hat s}\lsim 400$ GeV, the colliding photons
are polarized with $w^{+-}  \lsim  0.01$, $w^{++}=w^{--} \sim 0.4$, and
$w^{-+}  \sim  0.2$ (see the right panel of Fig.~\ref{fig:Lum_polarized}).
In contrast, when $\sqrt{\hat s}\gsim 680$ GeV, they become
$w^{+-}  \gsim  0.7$, $w^{++}=w^{--} \lsim 0.15$, and
$w^{-+}  \lsim  3\times 10^{-3}$.
Therefore, the high-$\sqrt{\hat s}$ region is expected to be well approximated
by the perfectly polarized scenario with $w^{+-}=1$,
whereas near the $2M_t$ threshold, the behavior likely falls
between the unpolarized case and the perfectly polarized scenario
with $w^{\pm\pm}=1$.

\begin{figure}[t!]
%\begin{figure}[h!]
%\begin{figure}[b!]
%\vspace{-0.5cm}
\begin{center}
\includegraphics[width=9.9cm,height=9.9cm]{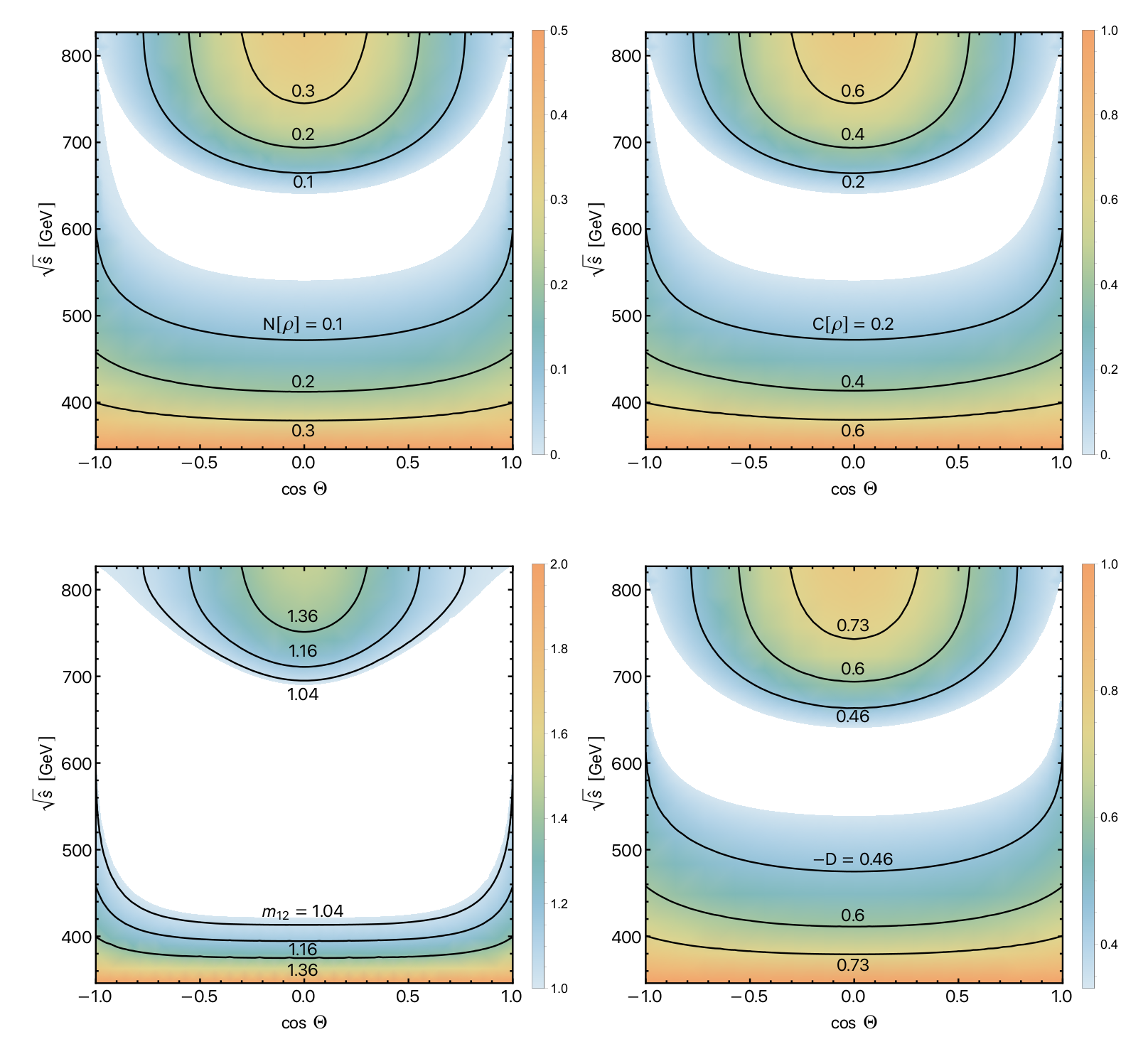}\hskip -0.5cm
\includegraphics[width=4.5cm,height=9.9cm]{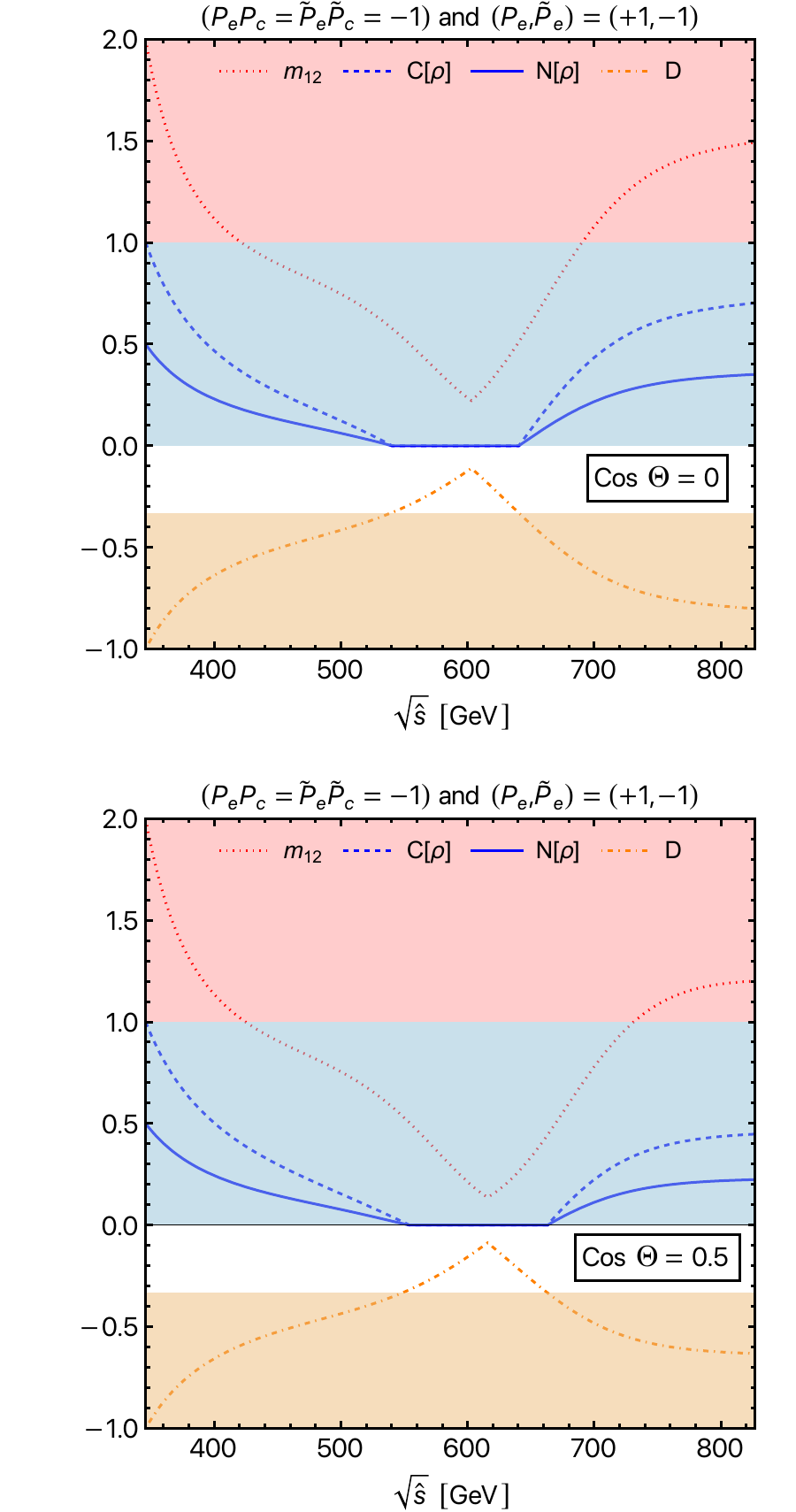}
\end{center}
\vspace{-0.5cm}
\caption{\it
{\bf Quantum entanglement with \boldmath{$\sqrt s= 1$} TeV and
\boldmath{$P_e=-\tilde P_e=+1$}}:
Same as in Fig.~\ref{fig:QE_500GeV} but for
$\sqrt s= 1$ TeV and $P_e=-P_c=-\tilde P_e=\tilde P_c=+1$.
}
\label{fig:QE_1TeV}
\end{figure}

Figure~\ref{fig:QE_1TeV} presents the contour regions for negativity
${\cal N}[\rho]$, concurrence ${\cal C}[\rho]$, the Bell nonlocality parameter
$m_{12}$, and the entanglement marker $D$, along with their dependence
on $\sqrt{\hat s}$.
A key difference from the perfectly polarized cases and the $\sqrt{s}=500$ GeV
case is that quantum entanglement and Bell inequality violation are not
observed throughout the entire $(\cos\Theta, \sqrt{\hat s})$ region.
Instead, along the $\cos\Theta=0$ line, quantum entanglement (${\cal N}[\rho]>0$
or ${\cal C}[\rho]>0$) is observed in the regions $\sqrt{\hat s}\lsim 540$ GeV
and $\sqrt{\hat s}\gsim 640$ GeV. Compared to the unpolarized case
(Fig.~\ref{fig:QE_unp}), where the corresponding regions are
$\sqrt{\hat s}\lsim 410$ GeV and $\sqrt{\hat s}\gsim 640$ GeV,
the entangled region is extended at lower $\sqrt{\hat s}$ in this polarized scenario.
Similarly, along the $\cos\Theta=0$ line, Bell inequality violation ($m_{12}>1$)
is observed in the regions $\sqrt{\hat s}\lsim 420$ GeV and
$\sqrt{\hat s}\gsim 690$ GeV. In contrast to the unpolarized case,
where Bell inequality violation occurs only for $\sqrt{\hat s}\lsim 370$ GeV
and $\sqrt{\hat s}\gsim 950$ GeV, the Bell nonlocality region is significantly
extended in both the low- and high-$\sqrt{\hat s}$ regions
in this polarized scenario.
It is worth noting in Fig.~\ref{fig:QE_1TeV} that the criterion
$D=\Delta_E/3<-1/3$ indeed provides a sufficient condition for entanglement.
This is evident by comparing the unshaded (white) separable region, where
${\cal N}[\rho]=0$ or ${\cal C}[\rho]=0$, with the region where $-D\leq 1/3$,
especially at large $|\cos\Theta|$.
We observe that the separable region shifts toward larger $\sqrt{\hat s}$ and shrinks as $\cos\Theta$ deviates from $0$.
%

%------------------------------------------------------------------
\subsection{Expected significances}
\label{sec:ExpectedSignificances}
To obtain the event rate considering the polarizations of the colliding photons,
we introduce the following dimensionless quantities:
\begin{eqnarray}
L^{\rm tot}\equiv\sum_{\lambda_1,\lambda_2} L^{\lambda_1\lambda_2}\,, \ \
L^{\rm avg}\equiv \frac{1}{4}L^{\rm tot}\,.
\end{eqnarray}
where, from Eqs.~\eqref{eq:weight_function} and \eqref{eq:w_L},
\begin{eqnarray}
L^{\lambda_1\lambda_2}=\frac{1}{{\cal L}_{ee}}
\frac{{\rm d}{\cal L}^{\lambda_1\lambda_2}_{\gamma\gamma}}{{\rm d}\sqrt\tau}
= w^{\lambda_1\lambda_2}L^{\rm tot}\,,
\end{eqnarray}
with 
\begin{eqnarray}
L^{\lambda_1\lambda_2} \ = \ 2\sqrt{\tau}\,
\left(\frac{1}{{\cal L}_{ee}}
\frac{{\rm d}{\cal L}^{\lambda_1\lambda_2}_{\gamma\gamma}}{{\rm d}\tau}\right)\,.
\end{eqnarray}
The $\gamma\gamma$ luminosity 
$\frac{1}{{\cal L}_{ee}}
\frac{{\rm d}{\cal L}^{\lambda_1\lambda_2}_{\gamma\gamma}}{{\rm d}\tau}$
for a specific combination of $\lambda_1$ and $\lambda_2$
is given by Eq.~\eqref{eq:Laa}.
Then the differential cross section is given by
\begin{eqnarray}
\frac{{\rm d}^2\sigma}{{\rm d}\sqrt\tau\,{\rm d}\cos\Theta} \ = \
\sum_{\lambda_1,\lambda_2}L^{\lambda_1\lambda_2}\,
\frac{{\rm d}\hat\sigma_{\lambda_1\lambda_2}}{{\rm d}\cos\Theta}\,,
\end{eqnarray}
where, dropping the sum and average over the spins 
of the colliding photons from Eq.~\eqref{eq:cx_0}, 
\begin{eqnarray}
\frac{{\rm d}\hat\sigma_{\lambda_1\lambda_2}}{{\rm d}\cos\Theta}\ = \
\frac{\beta\,N_c}{32\pi\hat s}\,\left|{\cal A}_C\right|^2\,
\sum_{\sigma,\bar\sigma}\left|\langle\sigma\bar\sigma;\lambda_1\lambda_2
\rangle\right|^2\,.
\end{eqnarray}

For the unpolarized case, 
$L^{\lambda_1\lambda_2}$ is independent of
$\lambda_1$ and $\lambda_2$ and we have
\begin{eqnarray}
L^{\lambda_1\lambda_2}=\frac{1}{{\cal L}_{ee}}
\frac{{\rm d}{\cal L}^{\rm unp}_{\gamma\gamma}}{{\rm d}\sqrt\tau}
\equiv L^{\rm unp} \,, \ \
L^{\rm tot}=4\,L^{\rm unp}\,, \ \
L^{\rm avg}=L^{\rm unp}\,,
\end{eqnarray}
where 
\begin{eqnarray}
L^{\rm unp} \ = \ 2\sqrt{\tau}\,
\left(\frac{1}{{\cal L}_{ee}}
\frac{{\rm d}{\cal L}^{\rm unp}_{\gamma\gamma}}{{\rm d}\tau}\right)\,,
\end{eqnarray}
with $\frac{1}{{\cal L}_{ee}}
\frac{{\rm d}{\cal L}^{\rm unp}_{\gamma\gamma}}{{\rm d}\tau}$
given by Eq.~\eqref{eq:Lunp}.
Then we have
\begin{eqnarray}
\frac{{\rm d}^2\sigma_0}{{\rm d}\sqrt\tau\,{\rm d}\cos\Theta}  &= &
\sum_{\lambda_1,\lambda_2}L^{\lambda_1\lambda_2}\left[
\frac{\beta\,N_c}{32\pi\hat s}\,\left|{\cal A}_C\right|^2\,
\sum_{\sigma,\bar\sigma}\left|\langle\sigma\bar\sigma;\lambda_1\lambda_2
\rangle\right|^2 \right]\nonumber \\ &=&
L^{\rm tot}\,
\frac{\beta\,N_c}{32\pi\hat s}\,\left|{\cal A}_C\right|^2\,
\left[\frac{1}{4}
\sum_{\lambda_1,\lambda_2,\sigma,\bar\sigma}
\left|\langle\sigma\bar\sigma;\lambda_1\lambda_2
\rangle\right|^2 \right]\nonumber \\ &=&
L^{\rm tot}\,
\frac{\beta\,N_c}{32\pi\hat s}\,\left|{\cal A}_C\right|^2\,
\left(\widehat C_1 \ + \ \widehat C_3\right)\,,
\end{eqnarray}
where we use $L^{\lambda_1\lambda_2}=L^{\rm unp}=L^{\rm avg}=L^{\rm tot}/4$ 
at the second 
equality. Note that the expressions are basically the same as
in Eq.~\eqref{eq:cx_unpol_0}.

When the colliding photons are polarized, we have
\begin{eqnarray}
\frac{{\rm d}^2\sigma}{{\rm d}\sqrt\tau\,{\rm d}\cos\Theta}  &= &
\sum_{\lambda_1,\lambda_2}L^{\lambda_1\lambda_2}\left[
\frac{\beta\,N_c}{32\pi\hat s}\,\left|{\cal A}_C\right|^2\,
\sum_{\sigma,\bar\sigma}\left|\langle\sigma\bar\sigma;\lambda_1\lambda_2
\rangle\right|^2 \right]\nonumber \\ &=&
L^{\rm tot}\,
\frac{\beta\,N_c}{32\pi\hat s}\,\left|{\cal A}_C\right|^2\,
\left[ \sum_{\sigma,\bar\sigma} \sum_{\lambda_1,\lambda_2}
w^{\lambda_1\lambda_2}\,
\left|\langle\sigma\bar\sigma;\lambda_1\lambda_2
\rangle\right|^2 \right]\nonumber \\ &=&
L^{\rm tot}\,
\frac{\beta\,N_c}{32\pi\hat s}\,\left|{\cal A}_C\right|^2\,
\left(\widehat C_1^w \ + \ \widehat C_3^w\right)\,,
\end{eqnarray}
where we use $L^{\lambda_1\lambda_2}=w^{\lambda_1\lambda_2}\,L^{\rm tot}$ 
at the second equality.
Then the event rate is given by
\begin{eqnarray}
\label{eq:event_rate}
\frac{{\rm d}^2  N}{{\rm d}\sqrt\tau\,{\rm d}\cos\Theta} =
\frac{\beta\,N_c}{32\pi\hat s}\,\left|{\cal A}_C\right|^2\,
\left[ \left({\cal L}_{ee}\,L^{\rm tot}\right)\,
\left(\widehat C_1^w + \widehat C_3^w\right)\right]\,,
\end{eqnarray}
where $\tau=\hat s/s$, $\beta = \sqrt{1 - 4M_t^2/\hat{s}}$, $N_c = 3$, and
${\cal A}_C = 8\pi\alpha Q_t^2/(1-\beta^2 \cos^2\Theta)$ with $Q_t=2/3$.
The total luminosity $L^{\rm tot}=4L^{\rm avg}$.
For $L^{\rm avg}$
with $P_eP_c=\tilde P_e\tilde P_c=-1$ and $x=4.8$,
see the dashed red line in any frame
in the left panel of Fig.~\ref{fig:Lum_polarized}.
The luminosity-weighted polarization coefficients
$\widehat C_1^w$ and $\widehat C_3^w$  are
\begin{eqnarray}
\widehat C_1^w & = & \sum_{\lambda_1,\lambda_2=\pm}
\ w^{\lambda_1\lambda_2}\, 
\left[|\langle ++;\lambda_1\lambda_2\rangle|^2
+ |\langle --;\lambda_1\lambda_2\rangle|^2\right]\,,
\nonumber \\
\widehat C_3^w & = & \sum_{\lambda_1,\lambda_2=\pm}
\ w^{\lambda_1\lambda_2}\, 
\left[|\langle +-;\lambda_1\lambda_2\rangle|^2
+ |\langle -+;\lambda_1\lambda_2\rangle|^2\right]\,.
\end{eqnarray}
where the weight functions are 
$w^{\lambda_1\lambda_2}={L^{\lambda_1\lambda_2}}/{L^{\rm tot}}$.
Taking $P_eP_c=\tilde P_e\tilde P_c=-1$ and $x=4.8$,
the helicity-dependent luminosities $L^{\lambda_1\lambda_2}$ and 
the weight functions $\omega^{\lambda_1\lambda_2}$ are shown in
the left and right panels of Fig.~\ref{fig:Lum_polarized}, respectively.
Finally, 
the reduced amplitudes $\langle \sigma\bar\sigma;\lambda_1\lambda_2\rangle$ 
are given in Table~\ref{tab:aatt_SM}. 
Once the event rate is given,
the polarization vectors~\eqref{eq:B_hat} and spin
correlations~\eqref{eq:C_hat} can be extracted from experimental data
by examining the angular distributions of the top decay products.
By integrating out the phase space of the decay products
while leaving over only one particle
in each decay of $t\to b f_u \bar f_d$ and
$\bar t\to \bar b \bar f_u f_d$,
one obtains the angular differential cross section of
\cite{Baumgart:2012ay}
\begin{equation}
\label{eq:ang_dist_0}
\frac{1}{\sigma}\frac{{\rm d}\sigma}{{\rm d}\Omega\,{\rm d}\overline\Omega}=
\frac{1}{(4\pi)^2}\left[
1+ \kappa\, {\bf B}\cdot\widehat{\bf q}
+ \bar\kappa\, \overline{\bf B}\cdot\widehat{\overline{\bf q}}
+ \kappa\bar\kappa\, \widehat{\bf q}\cdot{\bf C}\cdot\widehat{\overline{\bf q}}\,
\right]\,.
\end{equation}
In the $\{\hat n\,,\hat r\,,\hat k\}$ basis,
$\widehat{\bf q}=(s_\theta c_\phi, s_\theta s_\phi, c_\theta)$ and
$\widehat{\overline{\bf q}}=(s_{\bar\theta} c_{\bar\phi},
s_{\bar\theta} s_{\bar\phi}, c_{\bar\theta})$
in each one of the parent top and antitop  rest frames.
The vectors 
${\bf B}=(\widehat B_n^+,\widehat B_r^+,\widehat B_k^+)/\widehat A$ and 
$\overline{\bf B}=(\widehat B_n^-,\widehat B_r^-,\widehat B_k^-)/\widehat A$ are
nothing but the polarizations of top and antitop, respectively, and 
${\bf C}_{ij}=\widehat C_{ij}/\widehat A$ with $i,j=n,r,k$
is the spin correlation matrix,
see Eq.~\eqref{eq:SDM_hat}.
The analyzing powers for the top are given by
$\kappa_{\ell,d}=1$, $\kappa_{\nu,u}=-0.3$, and $\kappa_{b}=-0.4$ and
$\bar\kappa$ for the antitop by multiplying
by a minus sign or
$\bar\kappa_{\ell,d}=-1$, $\bar\kappa_{\nu,u}=+0.3$,
and $\bar\kappa_{b}=+0.4$.

If $\ell^\pm$ are chosen, for example, the angular distribution 
of top decay products becomes
\begin{equation}
\label{eq:ang_dist_ell}
\frac{1}{\sigma}\frac{{\rm d}\sigma}{{\rm d}\Omega_+{\rm d}\Omega_-}=
\frac{1}{(4\pi)^2}\left[
1+ {\bf B}^+\cdot\widehat{\bf q}_+
- {\bf B}^-\cdot\widehat{\bf q}_-
-\widehat{\bf q}_+\cdot{\bf C}\cdot\widehat{\bf q}_- \right]\,,
\end{equation}
and one could explicitly show that
\begin{equation}
\langle\cos\varphi\rangle=\int
\cos\varphi\
\left(
\frac{1}{\sigma}\frac{{\rm d}\sigma}{{\rm d}\Omega_+{\rm d}\Omega_-}\right)
{{\rm d}\Omega_+{\rm d}\Omega_-} =
-\frac{1}{9}\big(C_{nn}+C_{rr}+C_{kk}\big) =
-\frac{1}{9}\,{\rm Tr}[{\bf C}]\,,
\end{equation}
where
\begin{equation}
\cos\varphi\equiv\hat{\bf q}_+\cdot\hat{\bf q}_-=
s_{\theta_+} c_{\phi_+} s_{\theta_-} c_{\phi_-} +
s_{\theta_+} s_{\phi_+} s_{\theta_-} s_{\phi_-}+
c_{\theta_+} c_{\theta_-}
\end{equation}
with, for example,
\begin{eqnarray}
\langle c_{\theta_+} c_{\theta_-}\rangle  &=&
%\frac{1}{2}\langle c_{\theta_++\theta_-}+c_{\theta_--\theta_-}\rangle  =
\int c_{\theta_+} c_{\theta_-}
\left(
\frac{1}{\sigma}\frac{{\rm d}\sigma}{{\rm d}\Omega_+{\rm d}\Omega_-}\right)
{{\rm d}\Omega_+{\rm d}\Omega_-} \nonumber \\ &=&
\int
c_{\theta_+} c_{\theta_-}
\left[
\frac{1}{(4\pi)^2}\,
c_{\theta_+} c_{\theta_-} (-C_{kk})
\right]
{\rm d}c_{\theta_+}{\rm d}\phi_+
{\rm d}c_{\theta_-}{\rm d}\phi_-
\nonumber \\ &=&
(-C_{kk})\, \frac{1}{(4\pi)^2}\,
\left(\frac{2}{3}\right)\,
\left(\frac{2}{3}\right)\, (2\pi)\, (2\pi)
= -\frac{1}{9}C_{kk}\,.
\end{eqnarray}

In the $(\cos\Theta,\sqrt{\hat s})$ phase space
under consideration, we generate Monte Carlo (MC) events 
according to the event rate \eqref{eq:event_rate}.
For a given set of the values of $\cos\Theta$ and $\sqrt{\hat s}$,
the polarization vectors ${\bf B}$ and $\overline{\bf B}$ 
and the spin correlation matrix ${\bf C}$ are calculated to 
further generate the angular distribution of top decay products 
according to Eq.~\eqref{eq:ang_dist_0}.

Once the MC data set is ready, 
the components of ${\bf B}$, $\overline{\bf B}$,
and ${\bf C}$ are extracted from it by exploiting
the following averages over the angular distribution:
\begin{equation}
\langle\widehat{\bf q}\cdot \hat{\bf e}_i\rangle 
= \frac{\kappa}{3}\,B_i\,, \ \ \
\langle\widehat{\overline{\bf q}}\cdot \hat{\bf e}_j\rangle 
= \frac{\overline\kappa}{3}\,\overline{B}_j\,, \ \ \
\langle\widehat{\bf q}\cdot \hat{\bf e}_i\,
\widehat{\overline{\bf q}}\cdot \hat{\bf e}_j\rangle 
= \frac{\kappa\overline\kappa}{9}\,C_{ij}\,,
\end{equation}
where $i,j=n,r,k$ with $\hat{\bf e}_n=\hat n$, 
$\hat{\bf e}_r=\hat r$, and $\hat{\bf e}_k=\hat k$.
The extracted polarization vectors and the spin correlation 
matrix together with their statistical errors of
$\delta_{B_i}$, $\delta_{\overline{B}_j}$, and $\delta_{C_{ij}}$
are used to calculate 
the quantum entanglement quantifiers and estimate
the corresponding errors.
Though we do not construct an ensemble of 
ideally-well-defined genuine quantum states since
the $\{\hat n\,,\hat r\,,\hat k\}$ (helicity) basis 
changes event by event in the $(\cos\Theta,\sqrt{\hat s})$ phase space
under consideration,
one can still observe and measure quantum entanglement
and Bell inequality violation in this way
\cite{Cheng:2023qmz,Cheng:2024btk}.

For the practical purpose, we discretize the phase space into bins
with the bin size of $\Delta_{\sqrt{\hat s}}\times\Delta_{\cos\Theta}$.
In the $k$-th bin, 
the number of events is given by
\begin{equation}
n_k= \left.\frac{{\rm d}^2  N}{{\rm d}\sqrt\tau\,{\rm d}\cos\Theta}
\right|_{k{\rm -th}}\, 
\Delta_{\sqrt{\hat s}}\,\Delta_{\cos\Theta}\,\frac{1}{\sqrt{s}}\,,
\end{equation}
with  the event rate evaluated taking the midpoint values of
$\sqrt{\hat s}$ and $\cos\Theta$ of the $k$-th bin.
The components of the the polarization vectors ${\bf B}$ and $\overline{\bf B}$ 
and the spin correlation matrix ${\bf C}$ in the $k$-th bin
are also calculated using the midpoint values.
In each bin, we construct the per-bin concurrence 
${\cal C}^{(k)}[\rho^{(k)}]$ and
the per-bin Bell nonlocality parameter $m^{(k)}_{12}$ and make the
{\it conservative} estimation of their per-bin uncertainties as follows:
\begin{eqnarray}
\delta^{(k)}_{\cal C}&=&\sqrt{
 \sum_{i=n,r,k}\left(\delta^{(k)}_{B_i}\right)^2
+\sum_{j=n,r,k}\left(\delta^{(k)}_{\overline{B}_j}\right)^2
+\sum_{i,j=n,r,k}\left(\delta^{(k)}_{{C}_{ij}}\right)^2}\,,
\nonumber \\[2mm]
\delta^{(k)}_{m_{12}}&=&2\sqrt{m^{(k)}_{12}}\,\sqrt{
\sum_{i,j=n,r,k}\left(\delta^{(k)}_{{C}_{ij}}\right)^2}\,.
\end{eqnarray}
In the phase space where $m_{12}>1$~\eqref{eq:CHSH},
\footnote{Note that this is not an event-by-event cut. 
Instead, we are excluding the bins with $m^{(k)}_{12}<1$
from the phase space average. Incidentally,
also note that ${\cal C}[\rho]>0$~\eqref{eq:Concurrence} 
in the phase space where $m_{12}>1$.}
we estimate significances using the ratios 
\cite{Cheng:2024btk}
\begin{equation}
{\cal S}_{\cal C}=\frac{\langle\cal C\rangle}{\delta_{\cal C}}\,, \ \ \
{\cal S}_{m_{12}}=\frac{\langle m_{12}\rangle-1}{\delta_{m_{12}}}\,, 
\end{equation}
where event-count-weighted averages of the per-bin concurrence
${\cal C}^{(k)}[\rho^{(k)}]$ and 
per-bin Bell nonlocality parameter $m^{(k)}_{12}$ are given by
\begin{equation}
\langle{\cal C}\rangle=
\frac{\sum_k n_k\, {\cal C}^{(k)}[\rho^{(k)}]}{\sum_k n_k}\,, \ \ \
\langle m_{12}\rangle=
\frac{\sum_k n_k\, m^{(k)}_{12}}{\sum_k n_k}\,, \ \ \
\end{equation}
and the overall statistical uncertainties across all bins 
are obtained via the weighted averages of the per-bin ones:
\begin{equation}
\delta_C=\frac{1}{\sum_k n_k}
\sqrt{\sum_k n_k^2\left(\delta^{(k)}_C\right)^2} \,, \ \ \
\delta_{m_{12}}=\frac{1}{\sum_k n_k}
\sqrt{\sum_k n_k^2\left(\delta^{(k)}_{m_{12}}\right)^2}\,.
\end{equation}

Taking the bin size of $\Delta_{\sqrt{\hat s}}=3.85\,(16.35)$ GeV
for $\sqrt{s}=500\,(1000)$ GeV and $\Delta_{\cos\Theta}=0.1$ 
independently of $\sqrt{s}$,
we consider the following two benchmark scenarios
with $P_eP_c=\tilde P_e\tilde P_c=-1$ and $x=4.8$
based on the previous two subsections:
\footnote{We use 
${\rm BR}(W\to \ell\nu)=2\times 0.1086$ ($\ell=e,\mu$) and
${\rm BR}(W\to ~{\rm hadrons})=0.6741$
\cite{ParticleDataGroup:2024cfk}
with 
$\sigma(\gamma\gamma\to t\bar t)=1065$ fb ($\sqrt{s}=500$ GeV) and
$\sigma(\gamma\gamma\to t\bar t)=2345$ fb ($\sqrt{s}=1$ TeV).}
\begin{itemize}
\item{}$\sqrt{s}=500$ GeV with $P_e=\tilde P_e=1$:
$\sigma(\gamma\gamma\to t\bar t)
\times {\rm BR}(t\bar t \to \ell\ell\,,\ell j)=355$ fb
\item{}$\sqrt{s}=1$ TeV with $P_e=-\tilde P_e=1$:
$\sigma(\gamma\gamma\to t\bar t)
\times {\rm BR}(t\bar t \to \ell\ell\,,\ell j)=781$ fb
\end{itemize}
Note that both the dileptonic and semi-leptonic decay channels
of top quarks are taken into account and
ideal detector acceptance and 100\% reconstruction efficiency
are assumed.
%

%\begin{figure}[tb!]
%\begin{figure}[b!]
\begin{figure}[t!]
\centering
\includegraphics[width=0.48\linewidth]{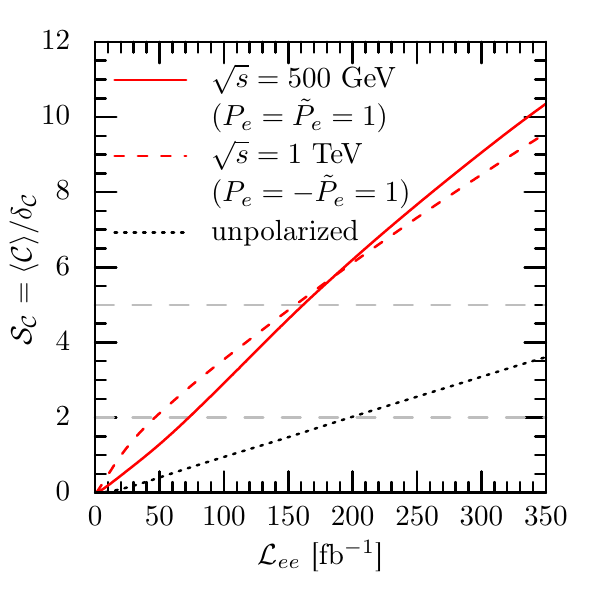}\hfill
\includegraphics[width=0.48\linewidth]{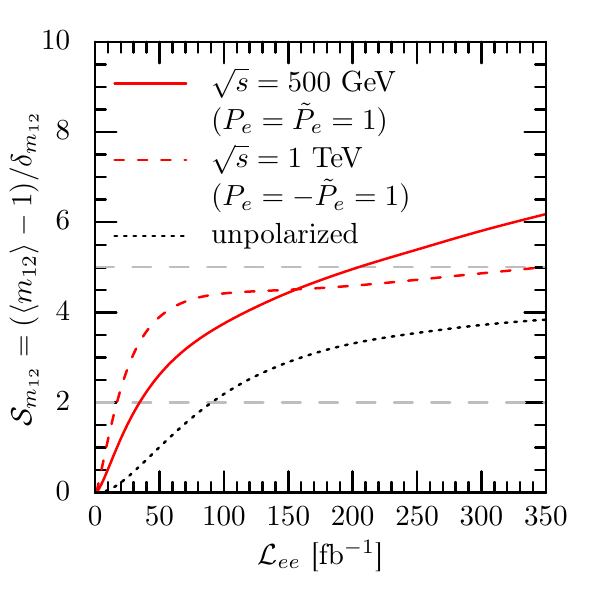}
\caption{
Statistical significances of 
the concurrence ${\cal S}_{\cal C}$ (left) and 
Bell nonlocality parameter ${\cal S}_{m_{12}}$ (right)
for the two benchmark scenarios  with
$\sqrt{s}=500$ GeV ($P_e=\tilde P_e=1$: solid red lines) and
$\sqrt{s}=1$ TeV ($P_e=-\tilde P_e=1$: dashed red lines).
We consider the full phase space for $\sqrt{s}=500$ GeV while
the phase-space bins with $m_{12}^{(k)} < 1$ are excluded from
the phase-space average for $\sqrt{s}=1$ TeV.
Both the dileptonic and semi-leptonic decay channels
of top quarks are taken into account and
ideal detector acceptance and 100\% reconstruction efficiency
are assumed.
The unpolarized beam scenario (dotted black lines)
corresponds to $\sqrt{s} = 500\text{ GeV}$
with $P_e = P_c = \tilde{P}_e = \tilde{P}_c = 0$.
}
\label{fig:significance}
\end{figure}
Figure~\ref{fig:significance}
displays the statistical significances ${\cal S}_{\cal C}$ (left)
and ${\cal S}_{m_{12}}$ (right). 
At $\sqrt{s}=500$ GeV, $m_{12} > 1$ throughout the entire 
accessible region of phase space (Fig.~\ref{fig:QE_500GeV}). In this
case, a violation of the Bell inequality could be observed with a
significance exceeding five standard deviations (${\cal S}_{m_{12}} >
5$) for integrated luminosities ${\cal L}_{ee} \gtrsim 200\text{
fb}^{-1}$. 
At $\sqrt{s}=1$ TeV, the condition $m_{12} > 1$ is not satisfied 
in the entire accessible region of phase space (Fig.~\ref{fig:QE_1TeV}).
For example, along the $\cos\Theta=0$ line, $m_{12} > 1$ 
where $\sqrt{\hat{s}} \lesssim 420$ GeV or 
$\sqrt{\hat{s}} \gtrsim 690$ GeV. 
In this case, a higher
luminosity of ${\cal L}_{ee} \gtrsim 350\text{ fb}^{-1}$ is required
to achieve ${\cal S}_{m_{12}} > 5$. 
Notably, quantum entanglement could be observed with a 
significance exceeding $5\sigma$ (${\cal S}_{\cal C} > 5$) 
with a bit and far lower luminosity of
${\cal L}_{ee}\gtrsim 170\text{ fb}^{-1}$ for 
$\sqrt{s}=500$ GeV and $\sqrt{s}=1$ TeV, respectively.
To provide a more realistic and decisive analysis, 
one would need to account for SM backgrounds, detector effects,
reconstruction efficiencies, systematic uncertainties, etc. However,
such considerations lie beyond the purpose and scope of the current study.

%

%------------------------------------------------------------------
\subsection{Comparisons}
\label{sec:Comparisons}
%
%

%\begin{figure}[t!]
%\begin{figure}[h!]
\begin{figure}[b!]
%\vspace{1.5cm}
\begin{center}
\includegraphics[width=15.0cm,height=6.5cm]{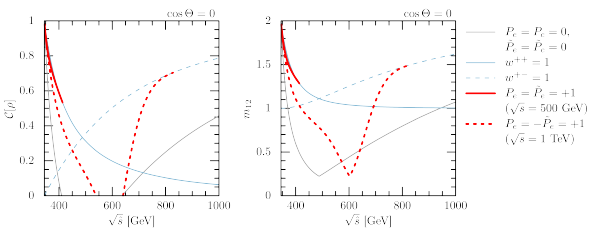}
\end{center}
\vspace{-0.5cm}
\caption{\it
{\bf Comparisons of quantum entanglement and Bell inequality violation}:
Concurrence ${\cal C}[\rho]$ (left) and
Bell nonlocality parameter $m_{12}$ (right)
as functions of $\sqrt{\hat s}$ at $\cos\Theta=0$
for unpolarized photons with $P_e=P_c=\tilde P_e=\tilde P_c=0$ (black solid),
perfectly polarized same-helicity photons (blue solid),
and perfectly polarized opposite-helicity photons (blue dashed).
Thick solid and dashed red curves correspond to
$P_e=-P_c=\tilde P_e=-\tilde P_c=+1$ ($\sqrt s= 500$ GeV) and
$P_e=-P_c=-\tilde P_e=+\tilde P_c=+1$ ($\sqrt s= 1$ TeV), respectively.
}
\label{fig:QE_comparisons}
\end{figure}

Before concluding, we compare the results obtained in our numerical analysis
regarding the impact of colliding-photon polarization on quantum entanglement
and Bell inequality violation.
Figure~\ref{fig:QE_comparisons} shows the concurrence ${\cal C}[\rho]$ (left) and
the Bell nonlocality parameter $m_{12}$ (right) as functions of $\sqrt{\hat s}$,
at $\cos\Theta=0$.
In the unpolarized case with $P_e=P_c=\tilde P_e=\tilde P_c=0$ (black solid line),
quantum entanglement is observed only near the $2M_t$ threshold and for
$\sqrt{\hat s}\gsim 640$ GeV, where ${\cal C}[\rho]>0$. The Bell nonlocality
range ($m_{12}>1$) is narrower.

When the colliding photons are perfectly polarized (blue solid and dashed lines),
quantum entanglement and the violation of the Bell inequality can be observed
across the entire range of $\sqrt{\hat{s}}$. Notably, the same-helicity and
opposite-helicity cases exhibit complementary behavior.
The same-helicity configuration demonstrates particular advantages near
the $2M_t$ threshold, while the opposite-helicity case shows superior
performance both across a broad range of $\sqrt{\hat{s}}$ and
specifically in the high-$\sqrt{\hat{s}}$ regime.
%
%
%
%
%In practice, the helicities of the colliding photons can be controlled
%through the use of polarized initial electrons and positrons combined
%with polarized laser beams.
To examine more realistic scenarios,
we consider two specific polarization configurations:
$P_e=-P_c=\tilde P_e=-\tilde P_c=+1$ and
$P_e=-P_c=-\tilde P_e=+\tilde P_c=+1$, evaluated at $\sqrt{s}= 500$ GeV
and $1$ TeV, respectively.
The first case with $\sqrt{s}= 500$ GeV is well approximated by the
perfectly polarized same-helicity case, as shown by
the thick solid red lines in Fig.~\ref{fig:QE_comparisons}.
Conversely, the second case with $\sqrt{s}= 1$ TeV closely matches
the perfectly polarized opposite-helicity case for
$\sqrt{\hat{s}}\gtrsim 750$ GeV, as illustrated by the thick
dashed red lines in Fig.~\ref{fig:QE_comparisons}.
Outside this high-energy region, the behavior falls intermediate
between the unpolarized and perfectly polarized cases.
In summary, our numerical analysis demonstrates that
the ability to control photon polarizations at a PLC substantially
improves the prospects for observing quantum entanglement and the violation of
the Bell inequality.

Last but not least, we compare our results with those obtained by studying
the process $e^+e^-\to t\bar t$ at future $e^+e^-$ colliders
taking account of the polarizations of the initial
electron and positron beams.
First of all, at LO in the SM, top-quark pair production at $e^+e^-$ colliders
proceeds via the $s$-channel $\gamma/Z$ exchanges and
it is reported that
quantum entanglement and Bell inequality violation 
are almost insensitive to the polarizations of the 
electron and positron beams~\cite{Cheng:2024btk,Altakach:2026fpl}.
This is contrasted to our finding that, at a photon collider,
quantum entanglement and Bell inequality violation strongly depend
on the polarizations of the colliding photons,
see the black and blue (solid and dashed) lines in Fig.~\ref{fig:QE_comparisons}.
Quantitatively, for $\sqrt{s}=500$ GeV,
the maximal values of concurrence ${\cal C}$ and
the Bell nonlocality parameter $m_{12}$ are about $0.29$ and $1.08$,%
\footnote{The quantity ${\cal B}$ considered 
in Refs.~\cite{Cheng:2024btk,Altakach:2026fpl} is related to
$m_{12}$ through ${\cal B}=2\sqrt{m_{12}}$.} respectively,
at $\cos\Theta\simeq -0.36$~\cite{Altakach:2026fpl}.%
\footnote{We also find the similar results on ${\cal B}$ 
with a selection cut of $\cos\Theta<0$ in Ref.~\cite{Cheng:2024btk}.}
Compared to our results ${\cal C}\gsim 0.5$ and $m_{12}\gsim 1.3$
at a photon collider,
see the thick solid red lines in Fig.~\ref{fig:QE_comparisons},
they are considerably weaker.
For $\sqrt{s}=1$ TeV, the maximal values of ${\cal C}$ and
$m_{12}$ are about $0.74$ and $1.54$, respectively,
at $\cos\Theta\simeq -0.46$~\cite{Altakach:2026fpl}.
They are larger compared to the case with $\sqrt{s}=500$ GeV
but note that the larger values up to  
${\cal C}^{\rm max}=1$ and $m_{12}^{\rm max}=2$ 
are attainable near the $t\bar t$ threshold at a photon collider,
see the thick dashed red lines in Fig.~\ref{fig:QE_comparisons}.
%

%------------------------------------------------------------------
\section{Conclusions and summary}
\label{sec:Conclusions}
A PLC, the two-photon collision mode of a linear $e^+e^-$ collider,
provides great opportunities to observe quantum entanglement
and the violation of the Bell inequality.
The polarizations of the colliding photons can be controlled through polarized
initial electrons and positrons combined with polarized laser beams.
This capability can
significantly enhance the magnitudes of the quantum entanglement quantifiers
such as
negativity ${\cal N}[\rho]$~\eqref{eq:Negativity},
concurrence ${\cal C}[\rho]$~\eqref{eq:Concurrence},
the Bell nonlocality parameter $m_{12}$~\eqref{eq:CHSH}, and
the entanglement marker $D$~\eqref{eq:D}.
To analyze the impact of colliding-photon polarizations
on the observation of quantum entanglement
through top-quark pair production at a PLC, $\gamma \gamma \to t \bar t$,
we have developed a formalism for the spin density matrix of a two-qubit
system constructed from the helicity amplitudes.

Once the spin density matrix is given,
% it is easy and straightforward to check
% the criteria for quantum entanglement and
% the violation of the Bell inequality.
the criteria for quantum entanglement and
the violation of the Bell inequality are readily evaluated.
The 16 polarization coefficients $\widehat C_i$
given in Appendix~\ref{sec:16Cs}
in terms of the reduced helicity amplitudes (Table~\ref{tab:aatt_SM})
determine the spin density matrix $\rho$ in Eq.~\eqref{eq:SDM_hat}
through the polarization vectors~\eqref{eq:B_hat} and spin
correlations~\eqref{eq:C_hat}
together with the normalization factor~\eqref{eq:A_hat}.
When the colliding photons are polarized, one can obtain
the spin density matrix by
using the luminosity-weighted polarization coefficients
$\widehat C_i^w$ in Eq.~\eqref{eq:C_weighted}.
We emphasize that our formalism
for the spin density matrix of a two-qubit system
is not only useful for the analysis of polarized-particle collisions
but also generically applicable to
any system composed of two spin-$1/2$ particles beyond LO
and/or beyond the SM.
It also
helps classify the polarization vectors and spin correlations of top quarks
according to their P, CP, and CP$\widetilde{\rm T}$ parities.

Through our numerical analysis, we clearly show that
the capability of controlling colliding-photon polarizations at a
PLC
significantly enhances the chances to observe
quantum entanglement and the violation of the Bell inequality
compared to the unpolarized case, as illustrated in Fig.~\ref{fig:QE_comparisons}.
%
% Pertinent to our numerical analysis, we suggest the following details
% as the points which need to be paid attention to:
Based on our numerical analysis, we highlight the following key points:
\begin{enumerate}
\item[$1$.]
In the unpolarized case with $P_e=P_c=\tilde P_e=\tilde P_c=0$,
quantum entanglement and Bell inequality violation
can be observed near the $2M_t$ threshold and only for
$\sqrt{\hat s}\gsim 640$ GeV and $\gsim 950$ GeV, respectively,
see Fig.~\ref{fig:QE_unp}.
\item[$2$.]
When the colliding photons are perfectly polarized with the same helicities,
quantum entanglement and Bell inequality violation
can be observed across the entire region of
$\cos\Theta$ and $\sqrt{\hat s}$, see Fig.~\ref{fig:QE_perfectPP}.
The quantum entanglement quantifiers do not depend on $\cos\Theta$, and
their magnitudes quickly {\it decrease} as $\sqrt{\hat s}$ increases.
\item[$3$.]
When the colliding photons are perfectly polarized with opposite helicities,
quantum entanglement and Bell inequality violation
can be observed again across the entire region of
$\cos\Theta$ and $\sqrt{\hat s}$, see Fig.~\ref{fig:QE_perfectPM}.
The magnitudes of the quantum entanglement quantifiers take their
largest values along the $\cos\Theta=0$ line, and they
{\it increase} as $\sqrt{\hat s}$ increases.
\item[$4$.]
For $\sqrt{s}=500$~GeV with $P_e=\tilde P_e=+1$,
$x=4.8$, and $P_eP_c=\tilde P_e\tilde P_c=-1$,
$\sqrt{\hat s}$ ranges from $2M_t$ to $410$~GeV.
The colliding photons are polarized with
$w^{++}  \gsim  0.7$, $w^{+-}=w^{-+} \lsim 0.15$,  and
$w^{--}  \lsim  3\times 10^{-3}$.
Quantum entanglement and Bell inequality violation
can be observed across the entire region of
$\cos\Theta$ and $\sqrt{\hat s}$, see Fig.~\ref{fig:QE_500GeV}.
This case is well approximated by
the perfectly polarized {\it same}-helicity configuration.
\item[$5$.]
For $\sqrt{s}=1$~TeV with $P_e=-\tilde P_e=+1$,
$x=4.8$, and $P_eP_c=\tilde P_e\tilde P_c=-1$,
$\sqrt{\hat s}$  ranges from $2M_t$ to $820$~GeV.
The magnitudes of the quantum entanglement quantifiers take their
largest values along the $\cos\Theta=0$ line, see Fig.~\ref{fig:QE_1TeV}.
Along the line,
quantum entanglement can be observed in the regions with
$\sqrt{\hat s}\lsim 540$ GeV and $\sqrt{\hat s}\gsim 640$ GeV, and
Bell inequality violation in the regions with
$\sqrt{\hat s}\lsim 420$ GeV and $\sqrt{\hat s}\gsim 690$ GeV.
This case is well approximated by the perfectly polarized {\it opposite}-helicity
configuration for $\sqrt{\hat{s}}\gsim 750$~GeV.
\item[$6$.]
Assuming ideal detector acceptance and 100\%
reconstruction efficiency, we find that a violation of the Bell
inequality could be observed with a significance exceeding five
standard deviations for integrated luminosities ${\cal L}_{ee}
\gtrsim 200\text{ fb}^{-1}$ and $350\text{ fb}^{-1}$ at
$\sqrt{s}=500$ GeV and $1$ TeV, respectively,
see Fig.~\ref{fig:significance}.
\end{enumerate}
In conclusion, the polarization control available at a PLC
provides a powerful handle for maximizing quantum entanglement
and observing Bell inequality violation in top-quark pair production.
The amplitude-level formalism developed in this work, with its
process- and model-independent classification of the spin density
matrix by P, CP, and CP$\widetilde{\mathrm{T}}$ parities, constitutes
a systematic framework directly applicable to other processes and to
new-physics searches beyond the SM.

%

%
%
%\newpage
%
%\section*{Acknowledgment}
%%
\acknowledgments

D.W.K. and C.B.P. thank the Asia Pacific Center for Theoretical Physics
(APCTP), Pohang, Korea, for their hospitality during the Focus Program
(APCTP-2026-F01), from which this work greatly benefited.
The work of S.Y.C. was supported by the Basic Science Research Program
through the National Research Foundation (NRF) of Korea
(No.\ RS-2022-NR075766) funded by the Ministry of Education.
The work of D.W.K. was supported by the NRF grant funded by the Korea
government (Ministry of Science and ICT, MSIT)
(No.\ RS-2025-02653237 and No.\ RS-2025-00562917) and 
Global - Learning Academic research institution 
for Master’s$\cdot$PhD students, and Postdocs (LAMP) Program of the 
NRF grant funded by the Ministry of Education (No. RS-2024-00443714).
The work of J.S.L. was supported by the NRF grant funded by the Korea
government (No.\ RS-2025-23523535).
The work of C.B.P. was supported by the NRF grant funded by the Korea
government (MSIT) (No.\ RS-2023-00209974).
This work was also supported in part by the NRF of Korea under
Grants No.\ RS-2022-NR070836 and No.\ RS-2024-00442775.
%RS-2022-NR070836 (SRC) RS-2024-00442775 (IUEP.. maybe)

%\paragraph{Note added.} This is also a good position for notes added
%after the paper has been written.

%\section*{Appendices}

\def\theequation{\Alph{section}.\arabic{equation}}
\begin{appendix}
%
%------------------------------------------------------
\setcounter{equation}{0}
\section{Polarization coefficients \boldmath{$\widehat C_i$} for
 \boldmath{$i$} from \boldmath{$1$} to \boldmath{$16$}}
%\section{Appendix}
\label{sec:16Cs}
%------------------------------------------------------
%
%
This appendix provides the explicit expressions for the 16 polarization
coefficients, $\widehat C_i$ (where $i$ ranges from $1$ to $16$), in terms of the
reduced helicity amplitudes, $\langle \sigma\bar\sigma;\lambda_1\lambda_2\rangle$.
These amplitudes are defined in Eq.~\eqref{eq:HelAmp} and their values for
the process under consideration are listed in Table~\ref{tab:aatt_SM}.
The expressions are given as follows:
\begin{eqnarray}
 \widehat C_1[+++]&=&\frac{1}{4}\sum_{\lambda_1,\lambda_2=\pm}
          \left[|\langle ++;\lambda_1\lambda_2\rangle|^2
              + |\langle --;\lambda_1\lambda_2\rangle|^2\right], \nonumber \\
 \widehat C_2[---]&=&\frac{1}{4}\sum_{\lambda_1,\lambda_2=\pm}
          \left[|\langle ++;\lambda_1\lambda_2\rangle|^2
              - |\langle --;\lambda_1\lambda_2\rangle|^2\right], \nonumber\\
 \widehat C_3[+++]&=&\frac{1}{4}\sum_{\lambda_1,\lambda_2=\pm}
          \left[|\langle -+;\lambda_1\lambda_2\rangle|^2
              + |\langle +-;\lambda_1\lambda_2\rangle|^2\right], \nonumber \\
 \widehat C_4[-++]&=&-\frac{1}{4}\sum_{\lambda_1,\lambda_2=\pm}
          \left[|\langle -+;\lambda_1\lambda_2\rangle|^2
              - |\langle +-;\lambda_1\lambda_2\rangle|^2\right],
\end{eqnarray}
\begin{eqnarray}
 \widehat C_5[-++]&=&\frac{1}{4}\,\real\sum_{\lambda_1,\lambda_2=\pm}
\big(\langle ++;\lambda_1\lambda_2\rangle
    - \langle --;\lambda_1\lambda_2\rangle\big)
\big(\langle -+;\lambda_1\lambda_2\rangle
    - \langle +-;\lambda_1\lambda_2\rangle\big)^*,   \nonumber\\
 \widehat C_6[++-]&=&-\frac{1}{4}\,\imag\sum_{\lambda_1,\lambda_2=\pm}
\big(\langle ++;\lambda_1\lambda_2\rangle
    - \langle --;\lambda_1\lambda_2\rangle\big)
\big(\langle -+;\lambda_1\lambda_2\rangle
    + \langle +-;\lambda_1\lambda_2\rangle\big)^*,   \nonumber\\
 \widehat C_7[---]&=&\frac{1}{4}\,\real\sum_{\lambda_1,\lambda_2=\pm}
\big(\langle ++;\lambda_1\lambda_2\rangle
    + \langle --;\lambda_1\lambda_2\rangle\big)
\big(\langle -+;\lambda_1\lambda_2\rangle
    + \langle +-;\lambda_1\lambda_2\rangle\big)^*,   \nonumber\\
 \widehat C_8[+-+]&=&-\frac{1}{4}\,\imag\sum_{\lambda_1,\lambda_2=\pm}
\big(\langle ++;\lambda_1\lambda_2\rangle
    + \langle --;\lambda_1\lambda_2\rangle\big)
\big(\langle -+;\lambda_1\lambda_2\rangle
    - \langle +-;\lambda_1\lambda_2\rangle\big)^*,
\nonumber \\
\end{eqnarray}
\begin{eqnarray}
 \widehat C_9[+--]&=&\frac{1}{4}\,\real\sum_{\lambda_1,\lambda_2=\pm}
\big(\langle ++;\lambda_1\lambda_2\rangle
    + \langle --;\lambda_1\lambda_2\rangle\big)
\big(\langle -+;\lambda_1\lambda_2\rangle
    - \langle +-;\lambda_1\lambda_2\rangle\big)^*,   \nonumber\\
 \widehat C_{10}[--+]&=&-\frac{1}{4}\,\imag\sum_{\lambda_1,\lambda_2=\pm}
\big(\langle ++;\lambda_1\lambda_2\rangle
    + \langle --;\lambda_1\lambda_2\rangle\big)
\big(\langle -+;\lambda_1\lambda_2\rangle
    + \langle +-;\lambda_1\lambda_2\rangle\big)^*,   \nonumber\\
 \widehat C_{11}[+++]&=&\frac{1}{4}\,\real\sum_{\lambda_1,\lambda_2=\pm}
\big(\langle ++;\lambda_1\lambda_2\rangle
    - \langle --;\lambda_1\lambda_2\rangle\big)
\big(\langle -+;\lambda_1\lambda_2\rangle
    + \langle +-;\lambda_1\lambda_2\rangle\big)^*,   \nonumber\\
 \widehat C_{12}[-+-]&=&-\frac{1}{4}\,\imag\sum_{\lambda_1,\lambda_2=\pm}
\big(\langle ++;\lambda_1\lambda_2\rangle
    - \langle --;\lambda_1\lambda_2\rangle\big)
\big(\langle -+;\lambda_1\lambda_2\rangle
    - \langle +-;\lambda_1\lambda_2\rangle\big)^*,
\nonumber \\
\end{eqnarray}
\begin{eqnarray}
 \widehat C_{13}[+++]&=&-\frac{1}{2}\,\real\sum_{\lambda_1,\lambda_2=\pm}
\big[\langle ++;\lambda_1\lambda_2\rangle
      \langle --;\lambda_1\lambda_2\rangle^*\big],\nonumber \\
    \widehat C_{14}[--+]&=&\frac{1}{2}\,\imag\sum_{\lambda_1,\lambda_2=\pm}
\big[\langle ++;\lambda_1\lambda_2\rangle
      \langle --;\lambda_1\lambda_2\rangle^*\big],  \nonumber\\
 \widehat C_{15}[+++]&=&-\frac{1}{2}\,\real\sum_{\lambda_1,\lambda_2=\pm}
\big[\langle -+;\lambda_1\lambda_2\rangle
      \langle +-;\lambda_1\lambda_2\rangle^*\big],\nonumber \\
   \widehat C_{16}[-+-]&=&-\frac{1}{2}\,\imag\sum_{\lambda_1,\lambda_2=\pm}
\big[\langle -+;\lambda_1\lambda_2\rangle
      \langle +-;\lambda_1\lambda_2\rangle^*\big].
\end{eqnarray}
The entries within the square brackets represent the P, CP, and
CP$\widetilde{\rm T}$ parities associated with each polarization coefficient.

%------------------------------------------------------
%\setcounter{equation}{0}
%\section{Appendix ... 2nd}
%\label{sec:NAME}
%------------------------------------------------------

\end{appendix}

%\newpage
%%%%%%%%%%%%%%%%%%%%%%%%%%%%%%%%%%%%%%%%

%%%%%%%%%%%%%%%%%%%%%%%%%%%%%%%%%%%%%%%%
%
\end{document}